\documentclass[zpreprint,zbstpl,british]{zeus_K2.9paper}
\usepackage{epsfig}
\usepackage{./zeusdet_hepunits}
\def\citeCTD{{\cite{%
nim:a279:290,*npps:b32:181,*nim:a338:254%
}}\xspace}
\def\citeMVD{{\cite{%
nim:a581:656%
}}\xspace}
\def\citeSTT{{\cite{%
nim:a535:191%
}}\xspace}
\def\citeCAL{{\cite{%
nim:a309:77,*nim:a309:101,*nim:a321:356,*nim:a336:23%
}}\xspace}
\def\citeHES{{\cite{%
nim:a277:176%
}}\xspace}
\def\citeSixmT{{\cite{%
thesis:gosau:2007%
}}\xspace}
\def\citeBAC{{\cite{%
nim:a300:480%
}}\xspace}
\def\citePCAL{{\cite{%
desy-92-066,*zfp:c63:391,*acpp:b32:2025%
}}\xspace}
\def\citeSPECTRO{{\cite{%
nim:a565:572%
}}\xspace}

\newcommand{\ZcoosysfnB}{%
The ZEUS coordinate system is a right-handed Cartesian system, with the $Z$
axis pointing in the proton beam direction, referred to as the ``forward
direction'', and the $X$ axis pointing towards the centre of HERA.
The coordinate origin is at the nominal interaction point.\xspace}

\newcommand{\ETjet}{\ensuremath{E_{T}^{\text{jet}}}\xspace}
\newcommand{\pTjet}{\ensuremath{p_{T}^{\text{jet}}}\xspace}

\newcommand{\ETgam}{\ensuremath{E_{T}^{\gamma}}\xspace}
\newcommand{\etagam}{\ensuremath{\eta^{\gamma}}\xspace}
\newcommand{\etajet}{\ensuremath{\eta^{\text{jet}}}\xspace}

\newcommand{\fmax}{\ensuremath{f_{\text{max}}}\xspace}

\begin{document}
%
%
\prepnum{DESY--12--089}
\prepdate{June 2012}                     

\title{Measurement of isolated photons accompanied by jets in deep inelastic {\textit{\textbf ep}} scattering}
                  
\author{ZEUS Collaboration}
\date{}              
\draftversion{}      

\maketitle

%
%
\begin{abstract}\noindent
{ The production of isolated high-energy photons accompanied by jets
has been measured in deep inelastic $ep$ scattering with the ZEUS
detector at HERA, using an integrated luminosity of $326\,
\mathrm{pb}^{-1}$. Measurements were made for exchanged photon
virtualities, $Q^2$, in the range $10$ to $350\,
\mathrm{GeV}^2$. The photons were measured in the transverse-energy and
pseudorapidity ranges $4<\ETgam< 15$ GeV and $-0.7 <\etagam< 0.9$, and
the jets were measured in the transverse-energy and pseudorapidity
ranges $2.5 <\ETjet<35$ GeV and $-1.5<\etajet< 1.8$. Differential
cross sections are presented as functions of these quantities.
Perturbative QCD predictions give a reasonable description of the
shape of the measured cross sections over most of the kinematic range,
but the absolute normalisation is typically in disagreement by 20-30\%.  }
\end{abstract}

\thispagestyle{empty}
\clearpage

%
%
%
%


\newcommand{\address}{ }
\newcommand{\author}{ }
                                                   %
\begin{center}
{                      \Large  The ZEUS Collaboration              }
\end{center}

{\small


        {\raggedright
H.~Abramowicz$^{45, ai}$, 
I.~Abt$^{35}$, 
L.~Adamczyk$^{13}$, 
M.~Adamus$^{54}$, 
R.~Aggarwal$^{7, c}$, 
S.~Antonelli$^{4}$, 
P.~Antonioli$^{3}$, 
A.~Antonov$^{33}$, 
M.~Arneodo$^{50}$, 
O.~Arslan$^{5}$
% ...
        }

\newpage


\makebox[3em]{$^{1}$}
\begin{minipage}[t]{14cm}
{\it Argonne National Laboratory, Argonne, Illinois 60439-4815, USA}~$^{A}$

\end{minipage}\\
\makebox[3em]{$^{2}$}
\begin{minipage}[t]{14cm}
{\it Andrews University, Berrien Springs, Michigan 49104-0380, USA}

\end{minipage}\\
\makebox[3em]{$^{3}$}
\begin{minipage}[t]{14cm}
{\it INFN Bologna, Bologna, Italy}~$^{B}$

\end{minipage}\\
\makebox[3em]{$^{4}$}
\begin{minipage}[t]{14cm}
{\it University and INFN Bologna, Bologna, Italy}~$^{B}$

\end{minipage}\\
\makebox[3em]{$^{5}$}
\begin{minipage}[t]{14cm}
{\it Physikalisches Institut der Universit\"at Bonn,
Bonn, Germany}~$^{C}$

\end{minipage}\\
\makebox[3em]{$^{6}$}
\begin{minipage}[t]{14cm}
{\it H.H.~Wills Physics Laboratory, University of Bristol,
Bristol, United Kingdom}~$^{D}$

\end{minipage}\\
\makebox[3em]{$^{7}$}
\begin{minipage}[t]{14cm}
{\it Panjab University, Department of Physics, Chandigarh, India}

\end{minipage}\\
\makebox[3em]{$^{8}$}
\begin{minipage}[t]{14cm}
{\it Calabria University,
Physics Department and INFN, Cosenza, Italy}~$^{B}$

\end{minipage}\\
\makebox[3em]{$^{9}$}
\begin{minipage}[t]{14cm}
{\it Institute for Universe and Elementary Particles, Chonnam National University,\\
Kwangju, South Korea}

\end{minipage}\\
\makebox[3em]{$^{10}$}
\begin{minipage}[t]{14cm}
{\it Jabatan Fizik, Universiti Malaya, 50603 Kuala Lumpur, Malaysia}~$^{E}$

\end{minipage}\\
\makebox[3em]{$^{11}$}
\begin{minipage}[t]{14cm}
{\it Nevis Laboratories, Columbia University, Irvington on Hudson,
New York 10027, USA}~$^{F}$

\end{minipage}\\
\makebox[3em]{$^{12}$}
\begin{minipage}[t]{14cm}
{\it The Henryk Niewodniczanski Institute of Nuclear Physics, Polish Academy of \\
Sciences, Krakow, Poland}~$^{G}$

\end{minipage}\\
\makebox[3em]{$^{13}$}
\begin{minipage}[t]{14cm}
{\it AGH-University of Science and Technology, Faculty of Physics and Applied Computer
Science, Krakow, Poland}~$^{H}$

\end{minipage}\\
\makebox[3em]{$^{14}$}
\begin{minipage}[t]{14cm}
{\it Department of Physics, Jagellonian University, Cracow, Poland}

\end{minipage}\\
\makebox[3em]{$^{15}$}
\begin{minipage}[t]{14cm}
{\it Deutsches Elektronen-Synchrotron DESY, Hamburg, Germany}

\end{minipage}\\
\makebox[3em]{$^{16}$}
\begin{minipage}[t]{14cm}
{\it Deutsches Elektronen-Synchrotron DESY, Zeuthen, Germany}

\end{minipage}\\
\makebox[3em]{$^{17}$}
\begin{minipage}[t]{14cm}
{\it INFN Florence, Florence, Italy}~$^{B}$

\end{minipage}\\
\makebox[3em]{$^{18}$}
\begin{minipage}[t]{14cm}
{\it University and INFN Florence, Florence, Italy}~$^{B}$

\end{minipage}\\
\makebox[3em]{$^{19}$}
\begin{minipage}[t]{14cm}
{\it Fakult\"at f\"ur Physik der Universit\"at Freiburg i.Br.,
Freiburg i.Br., Germany}

\end{minipage}\\
\makebox[3em]{$^{20}$}
\begin{minipage}[t]{14cm}
{\it School of Physics and Astronomy, University of Glasgow,
Glasgow, United Kingdom}~$^{D}$

\end{minipage}\\
\makebox[3em]{$^{21}$}
\begin{minipage}[t]{14cm}
{\it Department of Engineering in Management and Finance, Univ. of
the Aegean, Chios, Greece}

\end{minipage}\\
\makebox[3em]{$^{22}$}
\begin{minipage}[t]{14cm}
{\it Hamburg University, Institute of Experimental Physics, Hamburg,
Germany}~$^{I}$

\end{minipage}\\
\makebox[3em]{$^{23}$}
\begin{minipage}[t]{14cm}
{\it Imperial College London, High Energy Nuclear Physics Group,
London, United Kingdom}~$^{D}$

\end{minipage}\\
\makebox[3em]{$^{24}$}
\begin{minipage}[t]{14cm}
{\it Institute of Particle and Nuclear Studies, KEK,
Tsukuba, Japan}~$^{J}$

\end{minipage}\\
\makebox[3em]{$^{25}$}
\begin{minipage}[t]{14cm}
{\it Institute of Physics and Technology of Ministry of Education and
Science of Kazakhstan, Almaty, Kazakhstan}

\end{minipage}\\
\makebox[3em]{$^{26}$}
\begin{minipage}[t]{14cm}
{\it Institute for Nuclear Research, National Academy of Sciences, Kyiv, Ukraine}

\end{minipage}\\
\makebox[3em]{$^{27}$}
\begin{minipage}[t]{14cm}
{\it Department of Nuclear Physics, National Taras Shevchenko University of Kyiv, Kyiv, Ukraine}

\end{minipage}\\
\makebox[3em]{$^{28}$}
\begin{minipage}[t]{14cm}
{\it Kyungpook National University, Center for High Energy Physics, Daegu,
South Korea}~$^{K}$

\end{minipage}\\
\makebox[3em]{$^{29}$}
\begin{minipage}[t]{14cm}
{\it Institut de Physique Nucl\'{e}aire, Universit\'{e} Catholique de Louvain, Louvain-la-Neuve,\\
Belgium}~$^{L}$

\end{minipage}\\
\makebox[3em]{$^{30}$}
\begin{minipage}[t]{14cm}
{\it Departamento de F\'{\i}sica Te\'orica, Universidad Aut\'onoma
de Madrid, Madrid, Spain}~$^{M}$

\end{minipage}\\
\makebox[3em]{$^{31}$}
\begin{minipage}[t]{14cm}
{\it Department of Physics, McGill University,
Montr\'eal, Qu\'ebec, Canada H3A 2T8}~$^{N}$

\end{minipage}\\
\makebox[3em]{$^{32}$}
\begin{minipage}[t]{14cm}
{\it Meiji Gakuin University, Faculty of General Education,
Yokohama, Japan}~$^{J}$

\end{minipage}\\
\makebox[3em]{$^{33}$}
\begin{minipage}[t]{14cm}
{\it Moscow Engineering Physics Institute, Moscow, Russia}~$^{O}$

\end{minipage}\\
\makebox[3em]{$^{34}$}
\begin{minipage}[t]{14cm}
{\it Lomonosov Moscow State University, Skobeltsyn Institute of Nuclear Physics,
Moscow, Russia}~$^{P}$

\end{minipage}\\
\makebox[3em]{$^{35}$}
\begin{minipage}[t]{14cm}
{\it Max-Planck-Institut f\"ur Physik, M\"unchen, Germany}

\end{minipage}\\
\makebox[3em]{$^{36}$}
\begin{minipage}[t]{14cm}
{\it NIKHEF and University of Amsterdam, Amsterdam, Netherlands}~$^{Q}$

\end{minipage}\\
\makebox[3em]{$^{37}$}
\begin{minipage}[t]{14cm}
{\it Physics Department, Ohio State University,
Columbus, Ohio 43210, USA}~$^{A}$

\end{minipage}\\
\makebox[3em]{$^{38}$}
\begin{minipage}[t]{14cm}
{\it Department of Physics, University of Oxford,
Oxford, United Kingdom}~$^{D}$

\end{minipage}\\
\makebox[3em]{$^{39}$}
\begin{minipage}[t]{14cm}
{\it INFN Padova, Padova, Italy}~$^{B}$

\end{minipage}\\
\makebox[3em]{$^{40}$}
\begin{minipage}[t]{14cm}
{\it Dipartimento di Fisica dell' Universit\`a and INFN,
Padova, Italy}~$^{B}$

\end{minipage}\\
\makebox[3em]{$^{41}$}
\begin{minipage}[t]{14cm}
{\it Department of Physics, Pennsylvania State University, University Park,\\
Pennsylvania 16802, USA}~$^{F}$

\end{minipage}\\
\makebox[3em]{$^{42}$}
\begin{minipage}[t]{14cm}
{\it Polytechnic University, Tokyo, Japan}~$^{J}$

\end{minipage}\\
\makebox[3em]{$^{43}$}
\begin{minipage}[t]{14cm}
{\it Dipartimento di Fisica, Universit\`a 'La Sapienza' and INFN,
Rome, Italy}~$^{B}$

\end{minipage}\\
\makebox[3em]{$^{44}$}
\begin{minipage}[t]{14cm}
{\it Rutherford Appleton Laboratory, Chilton, Didcot, Oxon,
United Kingdom}~$^{D}$

\end{minipage}\\
\makebox[3em]{$^{45}$}
\begin{minipage}[t]{14cm}
{\it Raymond and Beverly Sackler Faculty of Exact Sciences, School of Physics, \\
Tel Aviv University, Tel Aviv, Israel}~$^{R}$

\end{minipage}\\
\makebox[3em]{$^{46}$}
\begin{minipage}[t]{14cm}
{\it Department of Physics, Tokyo Institute of Technology,
Tokyo, Japan}~$^{J}$

\end{minipage}\\
\makebox[3em]{$^{47}$}
\begin{minipage}[t]{14cm}
{\it Department of Physics, University of Tokyo,
Tokyo, Japan}~$^{J}$

\end{minipage}\\
\makebox[3em]{$^{48}$}
\begin{minipage}[t]{14cm}
{\it Tokyo Metropolitan University, Department of Physics,
Tokyo, Japan}~$^{J}$

\end{minipage}\\
\makebox[3em]{$^{49}$}
\begin{minipage}[t]{14cm}
{\it Universit\`a di Torino and INFN, Torino, Italy}~$^{B}$

\end{minipage}\\
\makebox[3em]{$^{50}$}
\begin{minipage}[t]{14cm}
{\it Universit\`a del Piemonte Orientale, Novara, and INFN, Torino,
Italy}~$^{B}$

\end{minipage}\\
\makebox[3em]{$^{51}$}
\begin{minipage}[t]{14cm}
{\it Department of Physics, University of Toronto, Toronto, Ontario,
Canada M5S 1A7}~$^{N}$

\end{minipage}\\
\makebox[3em]{$^{52}$}
\begin{minipage}[t]{14cm}
{\it Physics and Astronomy Department, University College London,
London, United Kingdom}~$^{D}$

\end{minipage}\\
\makebox[3em]{$^{53}$}
\begin{minipage}[t]{14cm}
{\it Faculty of Physics, University of Warsaw, Warsaw, Poland}

\end{minipage}\\
\makebox[3em]{$^{54}$}
\begin{minipage}[t]{14cm}
{\it National Centre for Nuclear Research, Warsaw, Poland}

\end{minipage}\\
\makebox[3em]{$^{55}$}
\begin{minipage}[t]{14cm}
{\it Department of Particle Physics and Astrophysics, Weizmann
Institute, Rehovot, Israel}

\end{minipage}\\
\makebox[3em]{$^{56}$}
\begin{minipage}[t]{14cm}
{\it Department of Physics, University of Wisconsin, Madison,
Wisconsin 53706, USA}~$^{A}$

\end{minipage}\\
\makebox[3em]{$^{57}$}
\begin{minipage}[t]{14cm}
{\it Department of Physics, York University, Ontario, Canada M3J 1P3}~$^{N}$

\end{minipage}\\
\vspace{30em} \pagebreak[4]


\makebox[3ex]{$^{ A}$}
\begin{minipage}[t]{14cm}
 supported by the US Department of Energy\
\end{minipage}\\
\makebox[3ex]{$^{ B}$}
\begin{minipage}[t]{14cm}
 supported by the Italian National Institute for Nuclear Physics (INFN) \
\end{minipage}\\
\makebox[3ex]{$^{ C}$}
\begin{minipage}[t]{14cm}
 supported by the German Federal Ministry for Education and Research (BMBF), under
 contract No. 05 H09PDF\
\end{minipage}\\
\makebox[3ex]{$^{ D}$}
\begin{minipage}[t]{14cm}
 supported by the Science and Technology Facilities Council, UK\
\end{minipage}\\
\makebox[3ex]{$^{ E}$}
\begin{minipage}[t]{14cm}
 supported by an FRGS grant from the Malaysian government\
\end{minipage}\\
\makebox[3ex]{$^{ F}$}
\begin{minipage}[t]{14cm}
 supported by the US National Science Foundation. Any opinion,
 findings and conclusions or recommendations expressed in this material
 are those of the authors and do not necessarily reflect the views of the
 National Science Foundation.\
\end{minipage}\\
\makebox[3ex]{$^{ G}$}
\begin{minipage}[t]{14cm}
 supported by the Polish Ministry of Science and Higher Education as a scientific project No.
 DPN/N188/DESY/2009\
\end{minipage}\\
\makebox[3ex]{$^{ H}$}
\begin{minipage}[t]{14cm}
 supported by the Polish Ministry of Science and Higher Education and its grants
 for Scientific Research\
\end{minipage}\\
\makebox[3ex]{$^{ I}$}
\begin{minipage}[t]{14cm}
 supported by the German Federal Ministry for Education and Research (BMBF), under
 contract No. 05h09GUF, and the SFB 676 of the Deutsche Forschungsgemeinschaft (DFG) \
\end{minipage}\\
\makebox[3ex]{$^{ J}$}
\begin{minipage}[t]{14cm}
 supported by the Japanese Ministry of Education, Culture, Sports, Science and Technology
 (MEXT) and its grants for Scientific Research\
\end{minipage}\\
\makebox[3ex]{$^{ K}$}
\begin{minipage}[t]{14cm}
 supported by the Korean Ministry of Education and Korea Science and Engineering
 Foundation\
\end{minipage}\\
\makebox[3ex]{$^{ L}$}
\begin{minipage}[t]{14cm}
 supported by FNRS and its associated funds (IISN and FRIA) and by an Inter-University
 Attraction Poles Programme subsidised by the Belgian Federal Science Policy Office\
\end{minipage}\\
\makebox[3ex]{$^{ M}$}
\begin{minipage}[t]{14cm}
 supported by the Spanish Ministry of Education and Science through funds provided by
 CICYT\
\end{minipage}\\
\makebox[3ex]{$^{ N}$}
\begin{minipage}[t]{14cm}
 supported by the Natural Sciences and Engineering Research Council of Canada (NSERC) \
\end{minipage}\\
\makebox[3ex]{$^{ O}$}
\begin{minipage}[t]{14cm}
 partially supported by the German Federal Ministry for Education and Research (BMBF)\
\end{minipage}\\
\makebox[3ex]{$^{ P}$}
\begin{minipage}[t]{14cm}
 supported by RF Presidential grant N 4142.2010.2 for Leading Scientific Schools, by the
 Russian Ministry of Education and Science through its grant for Scientific Research on
 High Energy Physics and under contract No.02.740.11.0244 \
\end{minipage}\\
\makebox[3ex]{$^{ Q}$}
\begin{minipage}[t]{14cm}
 supported by the Netherlands Foundation for Research on Matter (FOM)\
\end{minipage}\\
\makebox[3ex]{$^{ R}$}
\begin{minipage}[t]{14cm}
 supported by the Israel Science Foundation\
\end{minipage}\\
\vspace{30em} \pagebreak[4]


\makebox[3ex]{$^{ a}$}
\begin{minipage}[t]{14cm}
now at University of Salerno, Italy\
\end{minipage}\\
\makebox[3ex]{$^{ b}$}
\begin{minipage}[t]{14cm}
now at Queen Mary University of London, United Kingdom\
\end{minipage}\\
\makebox[3ex]{$^{ c}$}
\begin{minipage}[t]{14cm}
also funded by Max Planck Institute for Physics, Munich, Germany\
\end{minipage}\\
\makebox[3ex]{$^{ d}$}
\begin{minipage}[t]{14cm}
also Senior Alexander von Humboldt Research Fellow at Hamburg University,
 Institute of Experimental Physics, Hamburg, Germany\
\end{minipage}\\
\makebox[3ex]{$^{ e}$}
\begin{minipage}[t]{14cm}
also at Cracow University of Technology, Faculty of Physics,
 Mathemathics and Applied Computer Science, Poland\
\end{minipage}\\
\makebox[3ex]{$^{ f}$}
\begin{minipage}[t]{14cm}
supported by the research grant No. 1 P03B 04529 (2005-2008)\
\end{minipage}\\
\makebox[3ex]{$^{ g}$}
\begin{minipage}[t]{14cm}
supported by the Polish National Science Centre, project No. DEC-2011/01/BST2/03643\
\end{minipage}\\
\makebox[3ex]{$^{ h}$}
\begin{minipage}[t]{14cm}
now at Rockefeller University, New York, NY
 10065, USA\
\end{minipage}\\
\makebox[3ex]{$^{ i}$}
\begin{minipage}[t]{14cm}
now at DESY group FS-CFEL-1\
\end{minipage}\\
\makebox[3ex]{$^{ j}$}
\begin{minipage}[t]{14cm}
now at Institute of High Energy Physics, Beijing, China\
\end{minipage}\\
\makebox[3ex]{$^{ k}$}
\begin{minipage}[t]{14cm}
now at DESY group FEB, Hamburg, Germany\
\end{minipage}\\
\makebox[3ex]{$^{ l}$}
\begin{minipage}[t]{14cm}
also at Moscow State University, Russia\
\end{minipage}\\
\makebox[3ex]{$^{ m}$}
\begin{minipage}[t]{14cm}
now at University of Liverpool, United Kingdom\
\end{minipage}\\
\makebox[3ex]{$^{ n}$}
\begin{minipage}[t]{14cm}
now at CERN, Geneva, Switzerland\
\end{minipage}\\
\makebox[3ex]{$^{ o}$}
\begin{minipage}[t]{14cm}
also affiliated with Universtiy College London, UK\
\end{minipage}\\
\makebox[3ex]{$^{ p}$}
\begin{minipage}[t]{14cm}
now at Goldman Sachs, London, UK\
\end{minipage}\\
\makebox[3ex]{$^{ q}$}
\begin{minipage}[t]{14cm}
also at Institute of Theoretical and Experimental Physics, Moscow, Russia\
\end{minipage}\\
\makebox[3ex]{$^{ r}$}
\begin{minipage}[t]{14cm}
also at FPACS, AGH-UST, Cracow, Poland\
\end{minipage}\\
\makebox[3ex]{$^{ s}$}
\begin{minipage}[t]{14cm}
partially supported by Warsaw University, Poland\
\end{minipage}\\
\makebox[3ex]{$^{ t}$}
\begin{minipage}[t]{14cm}
now at Istituto Nucleare di Fisica Nazionale (INFN), Pisa, Italy\
\end{minipage}\\
\makebox[3ex]{$^{ u}$}
\begin{minipage}[t]{14cm}
now at Haase Energie Technik AG, Neum\"unster, Germany\
\end{minipage}\\
\makebox[3ex]{$^{ v}$}
\begin{minipage}[t]{14cm}
now at Department of Physics, University of Bonn, Germany\
\end{minipage}\\
\makebox[3ex]{$^{ w}$}
\begin{minipage}[t]{14cm}
now at Biodiversit\"at und Klimaforschungszentrum (BiK-F), Frankfurt, Germany\
\end{minipage}\\
\makebox[3ex]{$^{ x}$}
\begin{minipage}[t]{14cm}
also affiliated with DESY, Germany\
\end{minipage}\\
\makebox[3ex]{$^{ y}$}
\begin{minipage}[t]{14cm}
also at University of Tokyo, Japan\
\end{minipage}\\
\makebox[3ex]{$^{ z}$}
\begin{minipage}[t]{14cm}
now at Kobe University, Japan\
\end{minipage}\\
\makebox[3ex]{$^{\dagger}$}
\begin{minipage}[t]{14cm}
 deceased \
\end{minipage}\\
\makebox[3ex]{$^{aa}$}
\begin{minipage}[t]{14cm}
supported by DESY, Germany\
\end{minipage}\\
\makebox[3ex]{$^{ab}$}
\begin{minipage}[t]{14cm}
member of National Technical University of Ukraine, Kyiv Polytechnic Institute,
 Kyiv, Ukraine\
\end{minipage}\\
\makebox[3ex]{$^{ac}$}
\begin{minipage}[t]{14cm}
member of National University of Kyiv - Mohyla Academy, Kyiv, Ukraine\
\end{minipage}\\
\makebox[3ex]{$^{ad}$}
\begin{minipage}[t]{14cm}
partly supported by the Russian Foundation for Basic Research, grant 11-02-91345-DFG\_a\
\end{minipage}\\
\makebox[3ex]{$^{ae}$}
\begin{minipage}[t]{14cm}
Alexander von Humboldt Professor; also at DESY and University of Oxford\
\end{minipage}\\
\makebox[3ex]{$^{af}$}
\begin{minipage}[t]{14cm}
STFC Advanced Fellow\
\end{minipage}\\
\makebox[3ex]{$^{ag}$}
\begin{minipage}[t]{14cm}
now at LNF, Frascati, Italy\
\end{minipage}\\
\makebox[3ex]{$^{ah}$}
\begin{minipage}[t]{14cm}
This material was based on work supported by the
 National Science Foundation, while working at the Foundation.\
\end{minipage}\\
\makebox[3ex]{$^{ai}$}
\begin{minipage}[t]{14cm}
also at Max Planck Institute for Physics, Munich, Germany, External Scientific Member\
\end{minipage}\\
\makebox[3ex]{$^{aj}$}
\begin{minipage}[t]{14cm}
now at Tokyo Metropolitan University, Japan\
\end{minipage}\\
\makebox[3ex]{$^{ak}$}
\begin{minipage}[t]{14cm}
now at Nihon Institute of Medical Science, Japan\
\end{minipage}\\
\makebox[3ex]{$^{al}$}
\begin{minipage}[t]{14cm}
now at Osaka University, Osaka, Japan\
\end{minipage}\\
\makebox[3ex]{$^{am}$}
\begin{minipage}[t]{14cm}
also at \L\'{o}d\'{z} University, Poland\
\end{minipage}\\
\makebox[3ex]{$^{an}$}
\begin{minipage}[t]{14cm}
member of \L\'{o}d\'{z} University, Poland\
\end{minipage}\\
\makebox[3ex]{$^{ao}$}
\begin{minipage}[t]{14cm}
now at Department of Physics, Stockholm University, Stockholm, Sweden\
\end{minipage}\\
\makebox[3ex]{$^{ap}$}
\begin{minipage}[t]{14cm}
also at Cardinal Stefan Wyszy\'nski University, Warsaw, Poland\
\end{minipage}\\

}


\clearpage
\pagenumbering{arabic}
%

\section{Introduction}
\label{sec-int}

Events in which an isolated high-energy photon is observed provide a
direct probe of the underlying partonic process in high-energy
collisions involving hadrons, since the emission of such photons is
unaffected by parton hadronisation.  Processes of this kind have been
studied in a number of fixed-target and hadron-collider experiments
\cite{zfp:c13:207,*zfp:c38:371,*pr:d48:5,*prl:73:2662,*prl:95:022003,*prl:84:2786,*plb:639:151}.
In $ep$ collisions at HERA, the ZEUS and H1 collaborations have
previously reported the production of isolated photons in
photoproduction
\cite{pl:b413:201,pl:b472:175,pl:b511:19,epj:c49:511,epj:c38:437}, in
which the exchanged photon is quasi-real, and also in deep inelastic
scattering (DIS)~\cite{pl:b595:86,epj:c54:371,pl:b687:16}, where the
virtuality $Q^2$ of the exchanged virtual photon is greater than 1
\GeV$^2$.  The analysis presented here follows a recent ZEUS inclusive
measurement~\cite{pl:b687:16} of isolated photons in DIS.

Figure~\ref{fig1} shows the lowest-order tree-level diagrams for
high-energy photon production in DIS. Photons radiated by an incoming
or outgoing quark are called ``prompt''; an additional class of
photons comprises those radiated from the incoming or outgoing lepton.
In this paper, the inclusive photon measurements in DIS by ZEUS are
extended to include the requirement of a hadronic jet.  By increasing
the ratio of the prompt photon contribution relative to the
lepton-radiated contributions, this measurement provides an improved
test of perturbative QCD (pQCD) in a kinematic region with two hard
scales, which are given by $Q$ and by \pTjet, the transverse momentum
of the jet or, equivalently, the momentum transfer in the QCD
scatter. In particular, the fraction of prompt processes is increased,
and a class of jetless non-pQCD processes is excluded in which a soft
photon radiated within the proton undergoes a hard scatter off the
incoming electron~\cite{epj:c39:155}.  Compared to a previous ZEUS
publication on this topic~\cite{pl:b595:86}, the kinematic reach
extends to lower values of $Q^2$ and to higher values of the photon
transverse energy, $E_T^{\gamma}$, and the statistical precision is
much improved owing to the availability of nearly three times the
integrated luminosity.

Leading-logarithm parton-shower Monte Carlo (MC) and perturbative QCD
predictions are compared to the measurements.  The cross sections for
isolated photon production in DIS have been calculated to order ${O}
{(\alpha^3\alpha_s)}$ by Gehrmann-De Ridder 
{\it et al.\ }(GKS)~\cite{np:b578:326,prl:96:132002,epj:c47:395}.  A calculation
based on the $k_T$ factorisation approach has been made by Baranov
{\it et al.\ }(BLZ)~\cite{pr:d81:094034}.

\section{Experimental set-up}
\label{sec-exp}
The measurements are based on a data sample corresponding to an
integrated luminosity of $326\pm6\,\mathrm{pb}^{-1}$, taken 
during the years 2004 to 2007 with the ZEUS detector at HERA. During
this period, HERA ran with an electron/positron beam energy of 27.5
\GeV\  and a proton beam energy of 920 \GeV. The sample is a sum of
$138\pm2 \,\mathrm{pb}^{-1}$ of $e^+p$ data and $188\pm 3
\,\mathrm{pb}^{-1}$ of $e^-p$ data\footnote{Hereafter `electron'
refers to both electrons and positrons unless otherwise stated.}.

A detailed description of the ZEUS detector can be found
elsewhere~\cite{zeus:1993:bluebook}. Charged particles were tracked in
the central tracking detector (CTD)~\citeCTD and a silicon micro
vertex detector (MVD)~\cite{nim:a581:656} which operated in a magnetic
field of $1.43$~T provided by a thin superconducting solenoid.
The high-resolution uranium--scintillator calorimeter (CAL)~\citeCAL
consisted of three parts: the forward (FCAL), the barrel (BCAL) and
the rear (RCAL) calorimeters. The BCAL covered the pseudorapidity range
--0.74 to 1.01 as seen from the nominal interaction point. The FCAL
and RCAL extended the range to --3.5 to 4.0.  The smallest subdivision of
the CAL was called a cell. The barrel electromagnetic calorimeter
(BEMC) cells had a pointing geometry aimed at the nominal interaction
point, with a cross section approximately $5\times20\,\mathrm{cm^2}$,
with the finer granularity in the $Z$-direction\footnote{\ZcoosysfnB}.  This
fine granularity allows the use of shower-shape distributions to
distinguish isolated photons from the products of neutral meson decays
such as $\pi^0 \rightarrow \gamma \gamma$.

The luminosity was measured using the Bethe--Heitler reaction $ep
\rightarrow e\gamma p$ by a luminosity detector which consisted of two
independent systems: a lead--scintillator calorimeter
\cite{desy-92-066,*zfp:c63:391,*acpp:b32:2025} and a magnetic
spectrometer~\cite{nim:a565:572}.

\section{Event selection and reconstruction}
\label{sec-selec}

 A three-level trigger system was used to select events online
\cite{zeus:1993:bluebook,uproc:chep:1992:222,nim:a580:1257} by requiring well isolated
electromagnetic deposits in the CAL.

Events were selected offline by requiring a scattered-electron
candidate, identified using a neural
network~\cite{nim:a365:508,*nim:a391:360}.  The candidates were
required to have a polar angle in the range \mbox{$\theta_e
>140^{\circ} $}, in order to have a good measurement in the RCAL.  To
ensure a well understood acceptance, the impact point ($X$,$Y$) of the
candidate on the surface of the RCAL was required to lie outside a
rectangular region ($\pm 14.8\,\mathrm{cm}$ in $X$ and $[-14.6,
+12.5]$ cm in $Y$) centred on the origin of coordinates.  The energy
of the candidate, $E'_e$, was required to be larger than
$10\,\mathrm{\GeV}$. The kinematic quantities $Q^2$ and $x$ were
reconstructed from the scattered electron as $Q^2=-(k-k')^2$ and
\mbox{$x=Q^2/(2P\cdot(k-k')),$} where $k$ ($k'$) is the four-momentum
of the incoming (outgoing) lepton and $P$ is the four-momentum of the
incoming proton.  The kinematic region $10 <Q^2< 350\,\mathrm{\GeV^2}$
was selected.

To reduce backgrounds from non-$ep$ collisions, events were required
to have a reconstructed vertex position, $Z_{\mathrm{vtx}}$, within
the range $|Z_{\mathrm{vtx}}|< 40\,\mathrm{cm}$ and to have $35 <
E-p_Z< 65\,\mathrm{\GeV}$, where $E-p_Z = \sum \limits_i E_i(1-\cos
\theta_i)$; $E_i$ is the energy of the $i$-th CAL cell, $\theta_i$ is
its polar angle and the sum runs over all cells~\cite{pl:b303:183}.
The latter cut also removes events with large initial-state radiation
and low-$Q^2$ (photoproduction) events.

Energy-flow objects
(EFOs)~\cite{epj:c1:81,*epj:c6:43,*briskin:phd:1998} were constructed
from calorimeter-cell clusters, associated with tracks when
appropriate.  Photon candidates were identified as trackless EFOs for
which at least $90\%$ of the reconstructed energy was measured in the
BEMC.  EFOs with wider electromagnetic showers than are typical for a
single photon were accepted to allow evaluation of backgrounds.  The
reconstructed transverse energy of the photon candidate,
$E_T^{\gamma}$, was required to lie within the range \mbox{$4
<E_T^{\gamma}<15\,\mathrm{\GeV\ }$} and the pseudorapidity,
$\eta^{\gamma}$, had to satisfy $-0.7 < \eta^{\gamma} < 0.9$. The
upper limit on the reconstructed transverse energy was selected to
ensure that the shower shapes from the hadronic background and the
photon signal remained distinguishable.

Each event was required to contain an electron, a photon candidate and
at least one accompanying jet. Jet reconstruction was performed on all
EFOs in the event, including the electron and photon candidates, using
the $k_T$ clustering algorithm~\cite{np:b406:187} in the $E$-scheme in
the longitudinally invariant inclusive mode~\cite{pr:d48:3160} with
the $R$ parameter set to 1.0.  The jets were required to have
transverse energy, $E_T^{\mathrm{jet}}$, above 2.5 \GeV\ and to lie
within the pseudorapidity, $\eta^{\mathrm{jet}}$, range $-1.5
<\eta^{\mathrm{jet}} < 1.8$. One of the jets found by this procedure
corresponds to or includes the photon candidate. An additional
accompanying jet was required; if more than one was found, that with
the highest $E_T^{\mathrm{jet}}$ was used.

To reduce the background from photons and neutral mesons within jets,
and from photons radiated from electrons or positrons, the photon
candidate was required to be isolated from the reconstructed tracks
and other hadronic activity.  The isolation from tracks was achieved
by demanding $\Delta R>0.2$, where $\Delta R = \sqrt{(\Delta \phi)^2 +
(\Delta\eta)^2}$ is the distance to the nearest reconstructed track
with momentum greater than $250\,\mathrm{MeV}$ in the $\eta \-- \phi$
plane, where $\phi$ is the azimuthal angle.  Isolation from other
hadronic activity was imposed by requiring that the photon candidate
possessed at least $90 \%$ of the total energy of the reconstructed
jet of which it formed a part.

A total of 6167 events were selected at this stage; the sample was
dominated by background events.  The largest source of background came from 
neutral current DIS events in which the scattered electron was detected
in the RCAL, and one or more neutral mesons such as $\pi^0$ and $\eta$,
decaying to photons, produced a photon candidate in the BEMC.

\section{Theory}
\label{sec:theory}

\begin{figure}[p]
\vfill
\begin{center}
\epsfig{file=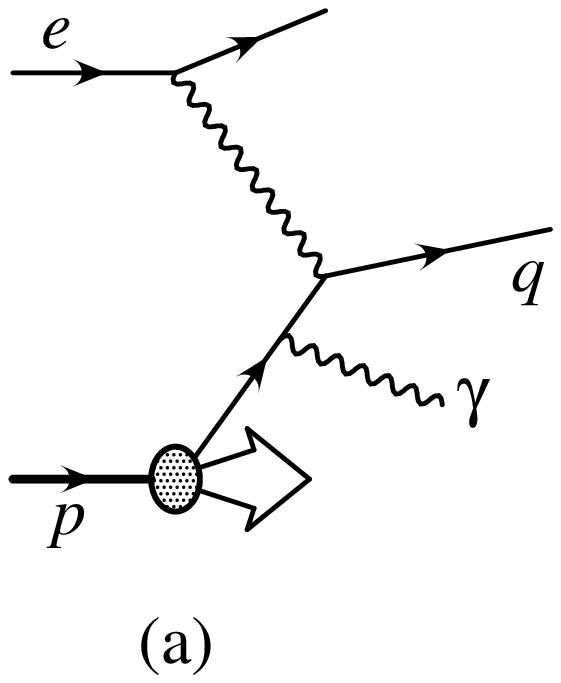,width=6cm}
\epsfig{file=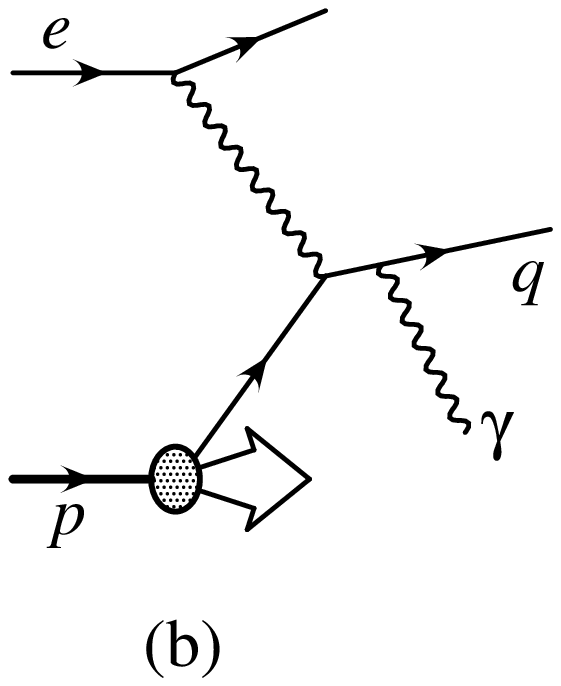,width=6cm}\\
\vspace{1.5cm}
\epsfig{file=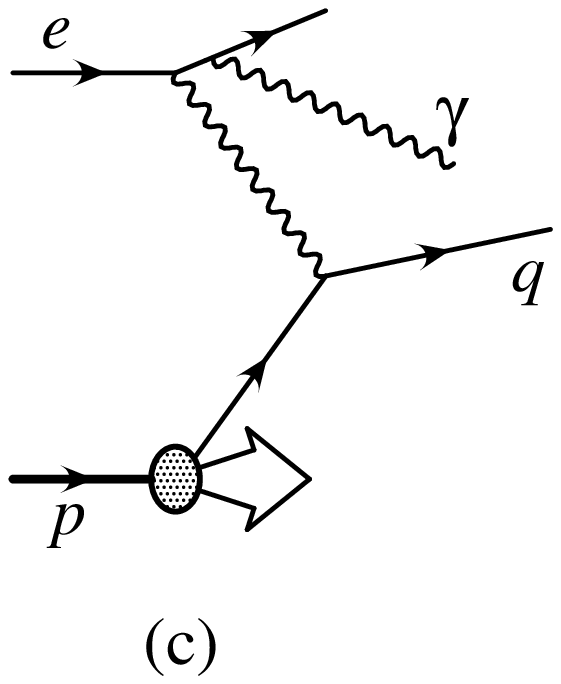,width=6cm}
\epsfig{file=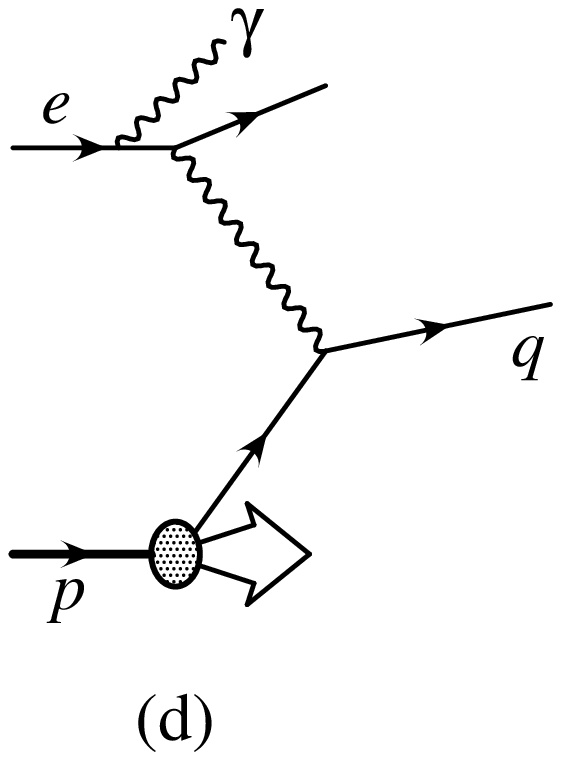,width=6cm}
\vspace{1.5cm}
\end{center}
\caption{\small Lowest-order tree-level diagrams for isolated photon 
production in 
$ep$ scattering.
(a) - (b):  quark radiative diagrams; (c) - (d): lepton radiative diagrams.
\label{fig1}}
\vfill
\end{figure}

Two theoretical predictions are compared to the measurements presented
in this paper. In the approach of
GKS~\cite{np:b578:326,prl:96:132002,epj:c47:395}, the contributions to
the scattering cross section for \mbox{$ep \rightarrow e\gamma X$} are
calculated at order $\alpha^3$, referred to here as LO, and
$\alpha^3\alpha_s$, referred to here as NLO, in the electromagnetic
and strong couplings. One of these contributions comes from the
radiation of a photon from the quark line (called QQ photons;
Fig.~\ref{fig1}a,b) and a second from the radiation from the lepton
line (called LL photons; Fig.~\ref{fig1}c,d). In addition to QQ and LL
photons, an interference term between photon emission from the lepton
and quark lines, called LQ photons by GKS, is present.  For the
kinematic region considered here, where the outgoing photon is well
separated from both outgoing electron and quark, the interference term
gives only a 3\% effect on the cross section. This effect is further
reduced to $\approx{1\%}$ when $e^+p$ and $e^-p$ data are combined as
the LQ term changes sign when $e^-$ is replaced by $e^+$.  The QQ
contribution includes photon emission at wide angles from the quark
as well as the leading $q\to q \gamma$ fragmentation term.

The GKS predictions use HERAPDF1.0 parton distribution functions for
the proton~\cite{jhep:01:109} and the BFG parton-photon fragmentation
functions~\cite{epj:c2:529}.  For their NLO calculation, the authors
quote an overall theoretical uncertainty of $(+4.3\%,-5.2\%)$ on
their integrated cross section, rising to approximately $\pm$10\% at
large negative jet rapidities. The uncertainty due to the choice of
proton parton distributions is typically much less than 5\%.  The
$k_T$ factorisation method used by BLZ~\cite{pr:d81:094034} takes into
account the photon radiation from the lepton as well as the quarks.
Unintegrated proton parton densities are used.  This procedure gives a
quark-radiated contribution that is enhanced relative to the
leading-order collinear approximations.  The uncertainties of up to
20\% in the calculation are due mainly to the procedure of selecting
jets from the evolution cascade in the factorisation approach.

In evaluating their predictions for the present data, both groups of
authors have incorporated the experimental selections and
photon-isolation procedure at the parton level.  Hadronisation
corrections were evaluated (see Section \ref{sec-mc}) to enable the
predictions to be compared to the experimental data which are
corrected to the hadron level.

\section{Monte Carlo event simulation}
\label{sec-mc}
Monte Carlo event samples were generated to evaluate the detector
acceptance and to provide signal and background distributions.
The program {\sc Pythia} 6.416~\cite{jhep:0605:026} was used to
simulate prompt-photon emission for the study of the
event-reconstruction efficiency. In {\sc Pythia}, this process is
simulated as a DIS process with additional photon radiation from the
quark line to account for QQ photons.  Radiation from the lepton is
not simulated.

The LL photons radiated at large angles from the incoming or outgoing
electron were simulated using the generator {\sc Djangoh}
6~\cite{cpc:81:381}, an interface to the MC program {\sc Heracles}
4.6.6~\cite{cpc:69:155}; higher-order QCD effects were included using
the colour dipole model of {\sc Ariadne} 4.12~\cite{cpc:71:15}.  
Hadronisation of the partonic final state was in each case performed
by {\sc Jetset} 7.4~\cite{cpc:39:347} using the Lund string
model~\cite{pr:97:31}. The small LQ contribution was neglected.

The main background to the QQ and LL photons came from photonic decays of
neutral mesons produced in general DIS processes. This background was
simulated using {\sc Djangoh} 6, within the same framework as the LL
events. This provided a realistic spectrum of single and multiple mesons 
with well modelled kinematic distributions. 

 The generated MC events were passed through the ZEUS detector and
trigger simulation programs based on {\sc Geant}
3.21~\cite{tech:cern-dd-ee-84-1}. They were reconstructed and analysed
by the same programs as the data.

Hadronisation corrections to the theory calculations were
evaluated using {\sc Pythia} and
{\sc Ariadne}, and typically lowered the theoretical prediction by
about 10\% with typical uncertainties of a few percent.  They were
calculated by running the same jet algorithm and event selections on
the generated partons and on the hadronised final state in the MC events.

\section{Extraction of the photon signal}

The event sample selected according to the criteria described in Section
\ref{sec-selec} was dominated by background; thus the photon signal was
extracted statistically following the approach used in previous ZEUS
analyses \cite{pl:b413:201,pl:b472:175,pl:b511:19,pl:b595:86,pl:b687:16}.

The photon signal was extracted from the background using the lateral
width of the BEMC energy-cluster comprising the photon candidate.
This was calculated as the variable $\langle\delta Z\rangle=
\sum \limits_i E_i|Z_i-Z_{\mathrm{cluster}}|$
{\large/}$( w_{\mathrm{cell}}\sum \limits_i E_i).$  Here, $Z_{i}$ is the $Z$
position of the centre of the $i$-th cell, $Z_{\mathrm{cluster}}$ is
the centroid of the EFO cluster, $w_{\mathrm{cell}}$ is the width of
the cell in the $Z$ direction, and $E_i$ is the energy recorded in the
cell. The sum runs over all BEMC cells in the EFO.

\begin{figure}[p]

\vfill
\begin{center}
\epsfig{file=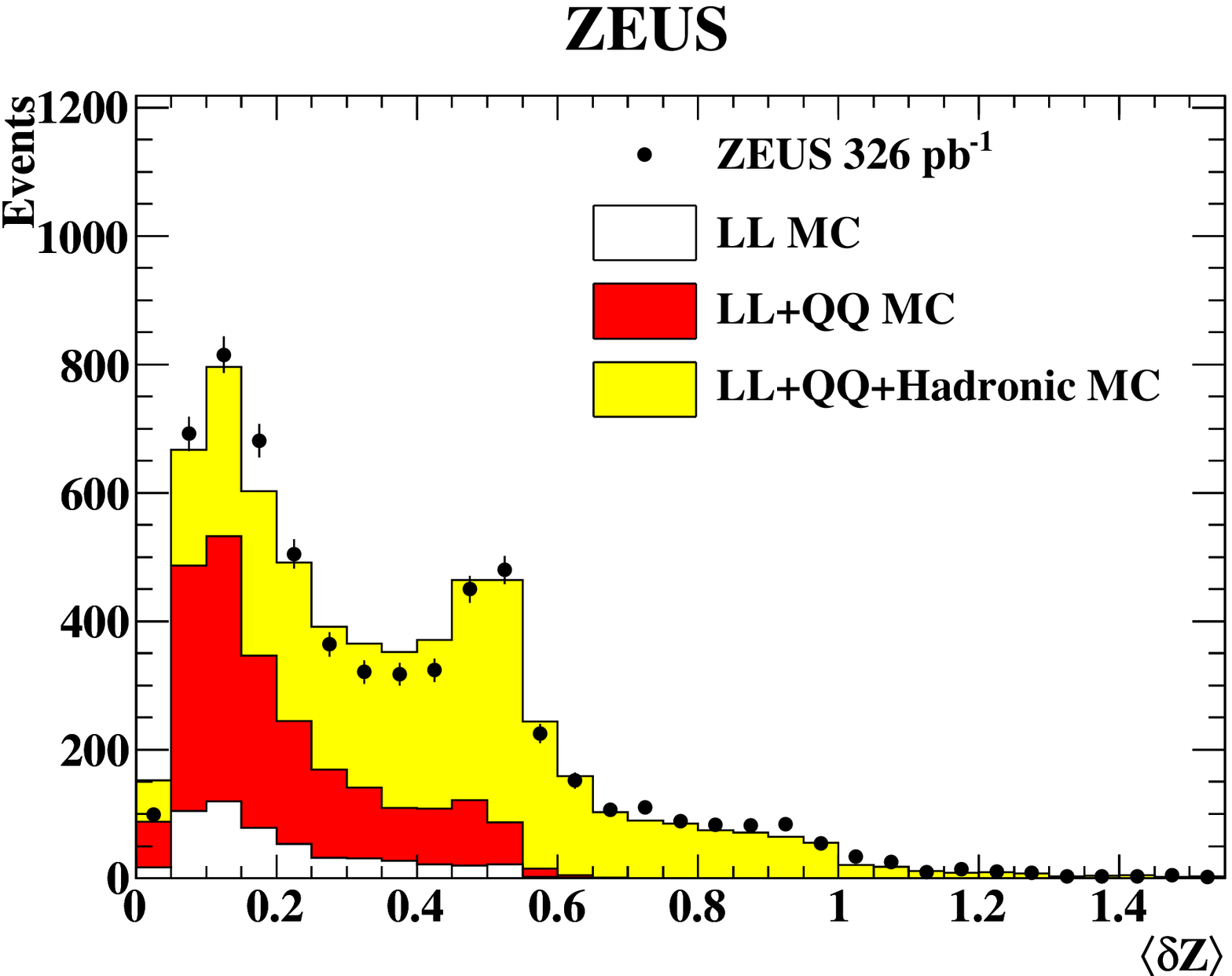,width=0.75\textwidth}\\[-0.01\textwidth]
\hspace*{0.01\textwidth}(a)\\[0.04\textwidth]
\epsfig{file=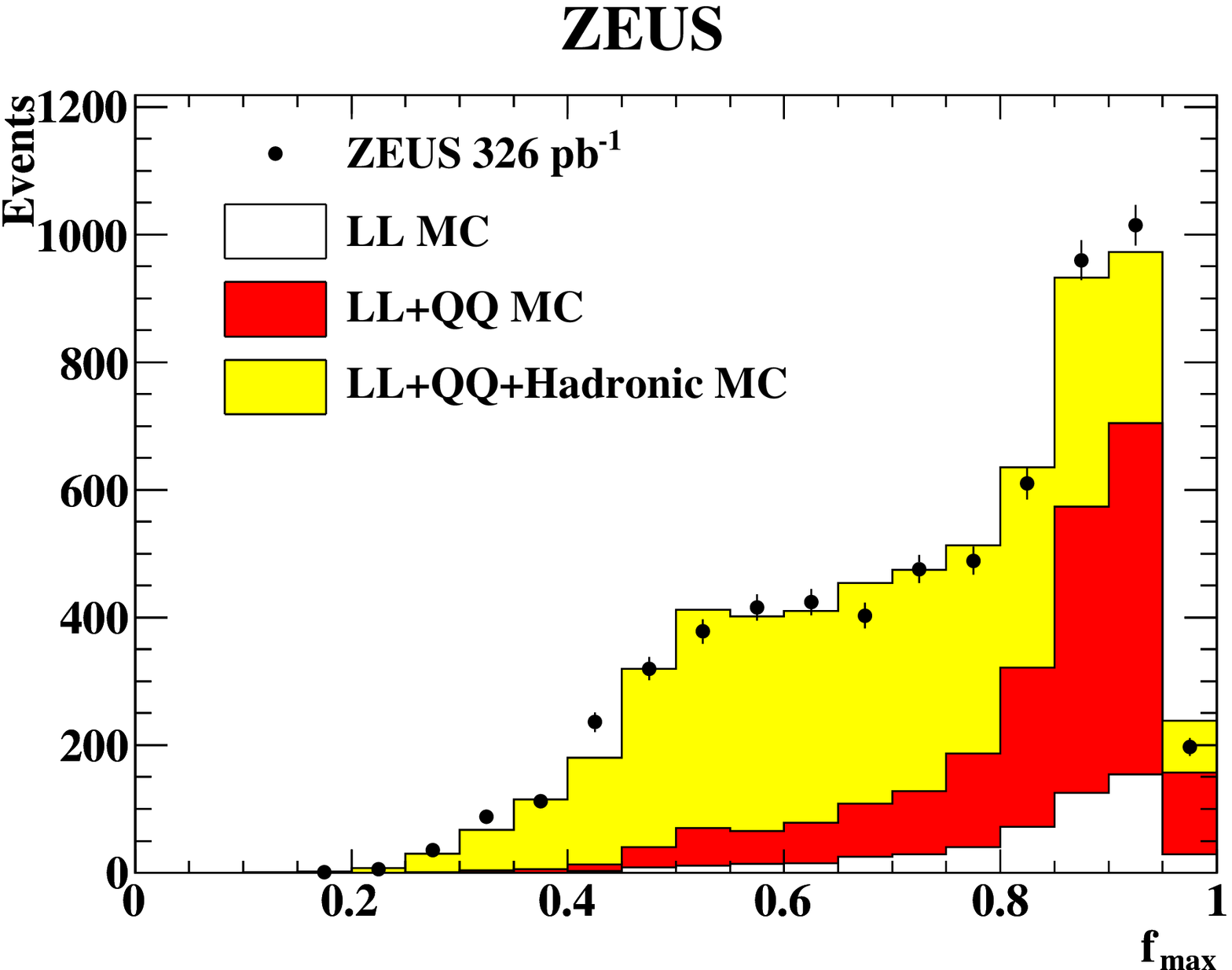,width=0.75\textwidth}\\[-0.01\textwidth]
\hspace*{0.01\textwidth}(b)\\[0.04\textwidth]
\end{center}
\caption{\small 
Distribution of (a) $\langle \delta Z \rangle$,  
(b) \fmax.
The error bars represent
the statistical uncertainties.  The light shaded histogram shows a fit
to the data of three components with fixed shapes as described in the
text. The dark shaded histogram represents the QQ component of the
fit, and the white histogram the LL component.
}
\label{fig:showers}
\vfill
\end{figure}

The global distribution of $\langle \delta Z \rangle$ in the data and
in the MC are shown in Fig.\,\ref{fig:showers}a. The MC distributions
in LL and QQ have been corrected using a comparison between the shapes
in $\langle \delta Z \rangle$ associated with the scattered electron
in MC simulation of DIS and in real data.  The
$\langle \delta Z \rangle$ distribution exhibits a double-peaked
structure with the first peak at $\approx 0.1$, associated with the
photon signal, and a second peak at $\approx 0.5$, dominated by the
$\pi^0\rightarrow\gamma \gamma$ background.

As a check, an alternative method was applied in which the quantity
\fmax was employed instead of $\langle\delta Z\rangle$, where \fmax is
the fraction of the photon-candidate shower contained in the BEMC cell
with the largest signal.  The results (Fig.\,\ref{fig:showers}b) were
consistent with the main analysis method and showed no significant
systematic difference.

The number of isolated-photon events contributing to the data is illustrated in 
Fig.\,\ref{fig:showers}a. It is determined for each cross-section bin 
by a $\chi^2$ fit to the $\langle \delta Z \rangle$ distribution in
the range $0<\langle \delta Z \rangle < 0.8$, using the LL and QQ
signal and background MC distributions as described in
Section~\ref{sec-mc}.  By treating the LL and QQ photons separately,
account is taken of their differing hadronic activity
(resulting in significantly different acceptances) and their differing
($\eta$, $E_{T}$) distributions (resulting in different bin
migrations due to finite measuring precision).

In performing the fit, the theoretically well determined LL
contribution was kept constant at its MC-predicted value and the other
components were varied.  Of the 6167 events selected, $2440\pm 60$
correspond to the extracted signal (LL and QQ). The scale factor
resulting from the global fit for the QQ photons in
Fig.\,\ref{fig:showers}a was 1.6; this factor was used for all the
plots comparing MC to data. The fitted global scale factor for the
hadronic background was 1.0.
The maximum value of $\chi^2/\mathrm{n.d.f.}$ of the fits in the cross section bins was 2.3 with an average of 1.5.

For a given observable $Y$, the production cross section was determined
using

\begin{equation}
\frac{d\sigma}{dY} = \frac{\mathcal{A}_{\mathrm{QQ}} \cdot N(\gamma_{\mathrm{QQ}})}{  
\mathcal{L} \cdot \Delta Y} + \frac{d\sigma^{\mathrm{MC}}_{\mathrm{LL}}}{dY} \nonumber ,
\end{equation}

where $N(\gamma_{\mathrm{QQ}})$ is the number of QQ photons extracted
from the fit, $\Delta Y$ is the bin width, $\mathcal{L}$ is the total
integrated luminosity, $\sigma^{\mathrm{MC}}_{\mathrm{LL}}$ is the
predicted cross section for LL photons from {\sc Djangoh}, and
$\mathcal{A}_{\mathrm{QQ}}$ is the acceptance correction for QQ
photons.  The value of $\mathcal{A}_{\mathrm{QQ}}$ was calculated
using Monte Carlo from the ratio of the number of events generated to
those reconstructed in a given bin.  It varied between 1.0 and 1.5
from bin to bin. To improve the representation of the data, and hence
the accuracy of the acceptance corrections, the Monte Carlo
predictions were reweighted.  This was done globally as a function of
$Q^2$ and of $\eta^{\gamma}$, and bin-by-bin as a function of
photon energy; the three reweighting factors were applied
multiplicatively.

\section{Systematic uncertainties}

The significant sources of systematic uncertainty were taken into
account as follows:

\begin{itemize}

\item the energy of the measured scattered electron was varied by its
known scale uncertainty of $\pm 2\%$~\cite{epj:c62:625}, causing
variations in the measured cross sections of up to $\pm 5\%$;

\item the energy of the photon candidate was similarly   
varied by $\pm 2\%$, causing variations in the measured cross sections
of up to $\pm 5\%$;

\item the modelling of the jets, and in particular the energy scale,
was first studied for jets with $\ETjet > 10$ \GeV\  by selecting ZEUS
DIS events having one jet of this type and no photon or other jets
with $\ETjet > 10$ \GeV. Using the scattered electron, and requiring
transverse-momentum balance, a prediction was made for the transverse
energy of the jet, which was compared to the values obtained 
in the data and in the MC events.  In this way, an uncertainty on the
energy scale of $\pm 1.5\%$ was established for these jets. For jets
with \ETjet in the range [2.5, 10] \GeV, DIS events were selected
containing one jet in this range and one jet with $\ETjet > 10$ \GeV.
Using the scattered electron and the well measured high-energy jet,
again requiring transverse-momentum balance, a prediction was made of the
lower jet \ETjet value, which was compared to the values obtained in
data and in MC.  In this way, the uncertainty on the jet energy scale was
evaluated as $\pm 4\%$ and $\pm 2.5\%$ in the energy ranges [2.5, 6]
and [6, 10] \GeV, respectively.  The resulting systematic uncertainty
on the cross section  was typically around $\pm2\%$, ranging to
$\pm10\%$ at the highest \ETjet values.

\end{itemize}

Since the photon and jet energy scales were calibrated relative to that of
the scattered electron, all three energy-scale uncertainties were treated as
correlated.  The three energy scales were simultaneously varied by the
uncertainties described above, and the resulting change in the cross sections
was taken as the overall systematic energy-scale uncertainty.
Further systematic uncertainties were evaluated as follows:

\begin{itemize}
\item  the dependence on the modelling of the hadronic background by 
{\sc Ariadne} was investigated by varying the upper limit for the
$\langle \delta Z\rangle$ fit in the range $[0.6, 1.0]$, giving
variations that were typically $\pm5\%$ increasing to +12\% and $-14\%$ in
the most forward $\eta^\gamma$ and highest-$x$ bins respectively;

\item  uncertainties in the acceptance due to the modelling by {\sc Pythia} were
accounted for by taking half of the change attributable to the reweighting 
as a systematic uncertainty; for most points the effect was small.

\end{itemize}
The background from photoproduction events at low $Q^2$ was found to
be negligible.  Other sources of systematic uncertainty were found to
be negligible and were ignored~\cite{forrest:phd:2009,pl:b687:16}: these included the
modelling of the $\Delta R$ cut, the track momentum cut, the cut on
$E-p_Z$, the $Z_{\text{vtx}}$ cut, the cut on the electromagnetic fraction of
the photon shower, and a variation of 5\% on the LL fraction.  These
were found to generate systematic effects of at most 1-2\% apart from
a 2.5\% effect in the highest-$x$ bin.

The major uncertainties were treated as symmetric and added in
quadrature.  The common uncertainty of $1.8\% $ on the
luminosity measurement was not included in the tables and figures.

\section{Results}

\label{sec:results}

Differential cross sections in DIS for the production of an isolated
photon and at least one additional jet, $ep\rightarrow e'
\gamma+\mathrm{jet}$, were measured in the kinematic region defined
by $10 < Q^2< 350 \, \mathrm{\GeV}^2$, $E'_e > 10\,\mathrm{\GeV}$,
$\theta_e > 140^{\circ}$, $-0.7< \eta^\gamma < 0.9$, $4 <
E_T^{\gamma}< 15\,\mathrm{\GeV} $, $\ETjet > 2.5$ \GeV\ and $-1.5
<\etajet< 1.8$ in the laboratory frame. The jets were formed according
to the $k_T$-clustering algorithm with the $R$ parameter set to 1.0,
and photon isolation was imposed such that at least $90 \%$ of the
energy of the jet-like object containing the photon belongs to the
photon.  No track with momentum greater than 250 MeV was
allowed within a cone around the photon of radius 0.2 in $\eta,\,
\phi$.

\begin{figure}[p]
\vfill
\begin{center}
\mbox{
\hspace*{-0.05\textwidth}
\epsfig{file=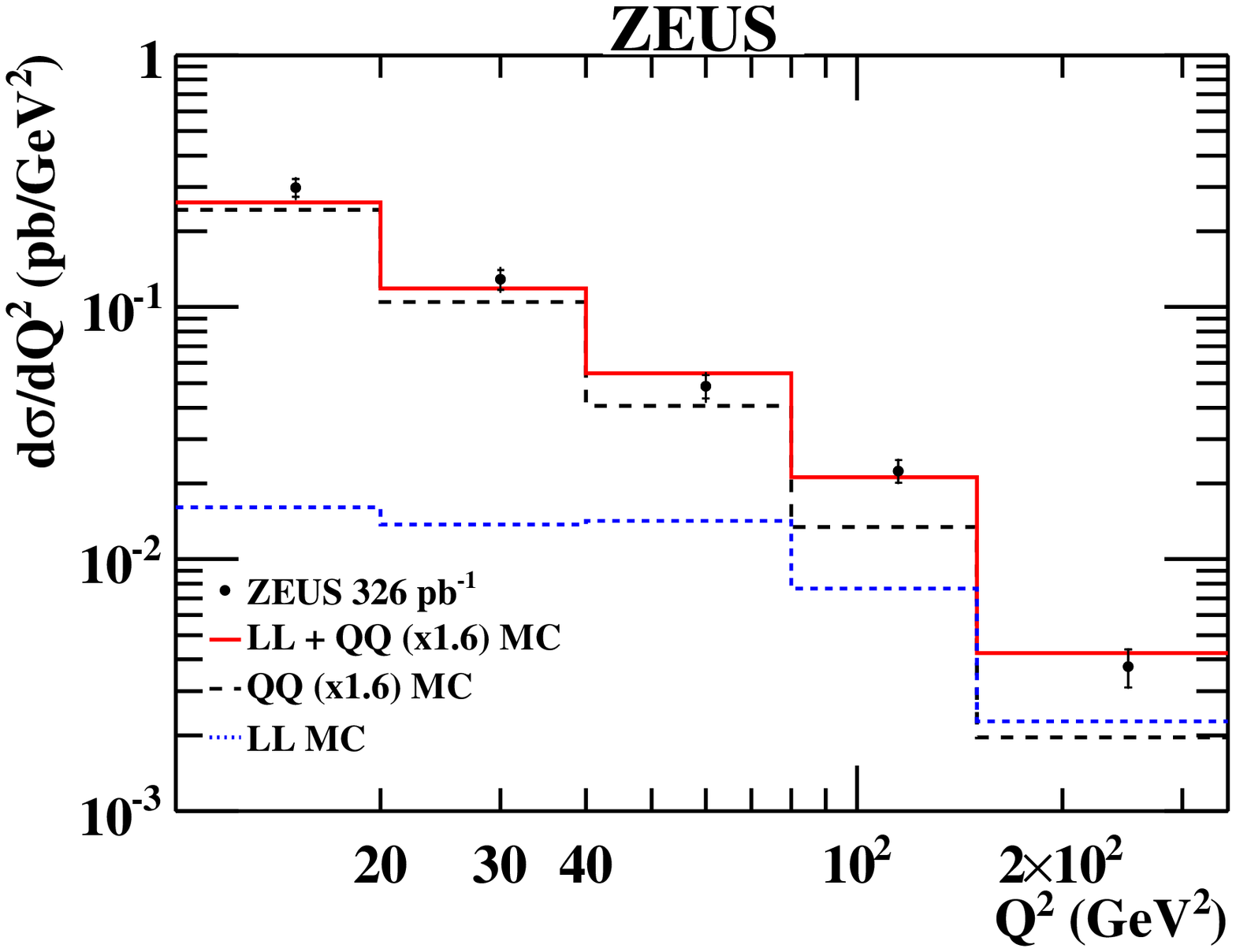,width=0.54\textwidth}
\hspace*{-0.04\textwidth}
\epsfig{file=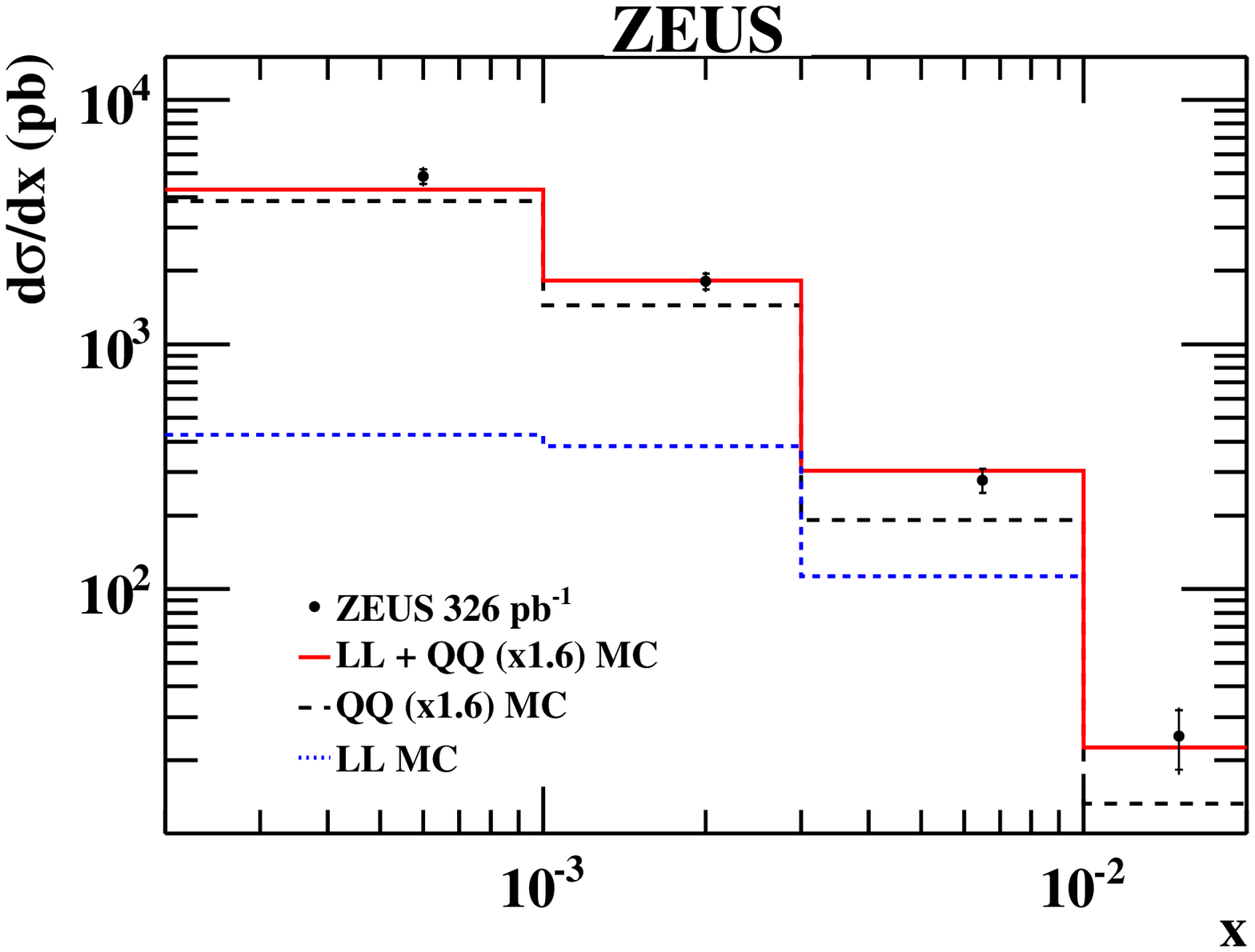,width=0.54\textwidth}
}\\[-0.04\textwidth]
\hspace*{0.1\textwidth}(a)\hspace*{0.48\textwidth}(b)\\[0.04\textwidth]
\mbox{
\hspace*{-0.05\textwidth}
\epsfig{file=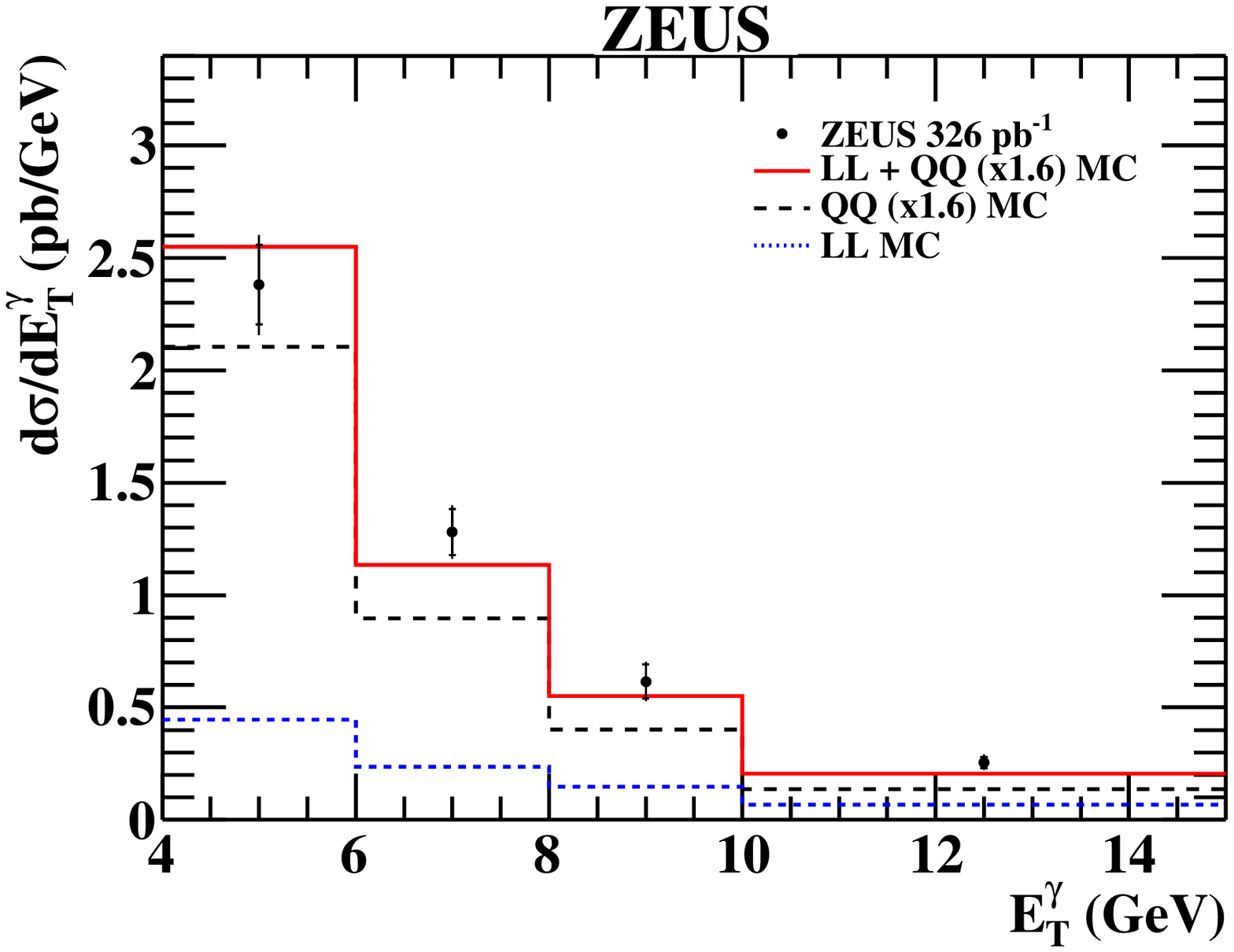,width=0.54\textwidth}
\hspace*{-0.04\textwidth}
\epsfig{file=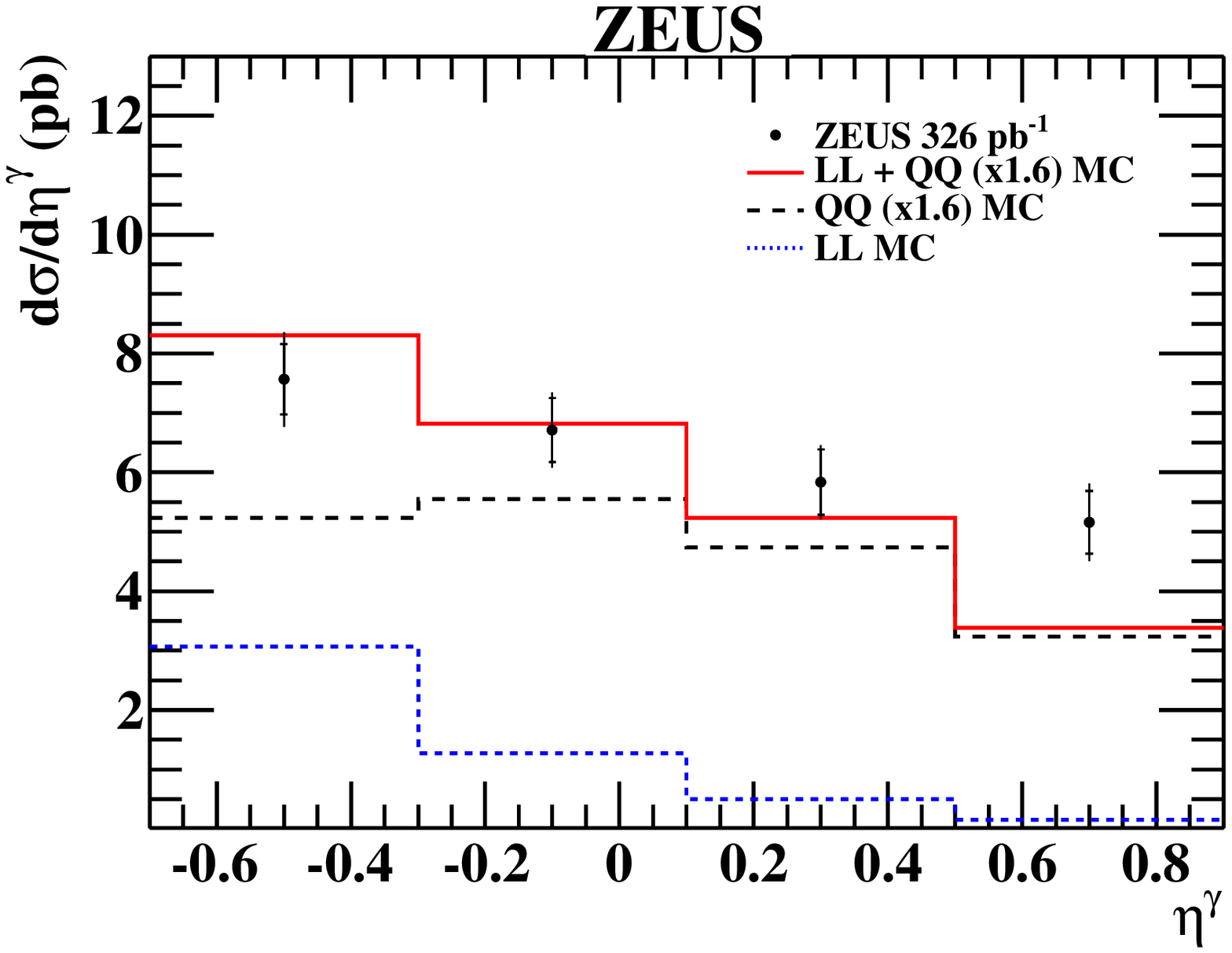,width=0.54\textwidth}
}\\[-0.04\textwidth]
\hspace*{0.1\textwidth}(c)\hspace*{0.48\textwidth}(d)\\[0.04\textwidth]
\mbox{
\hspace*{-0.05\textwidth}
\epsfig{file=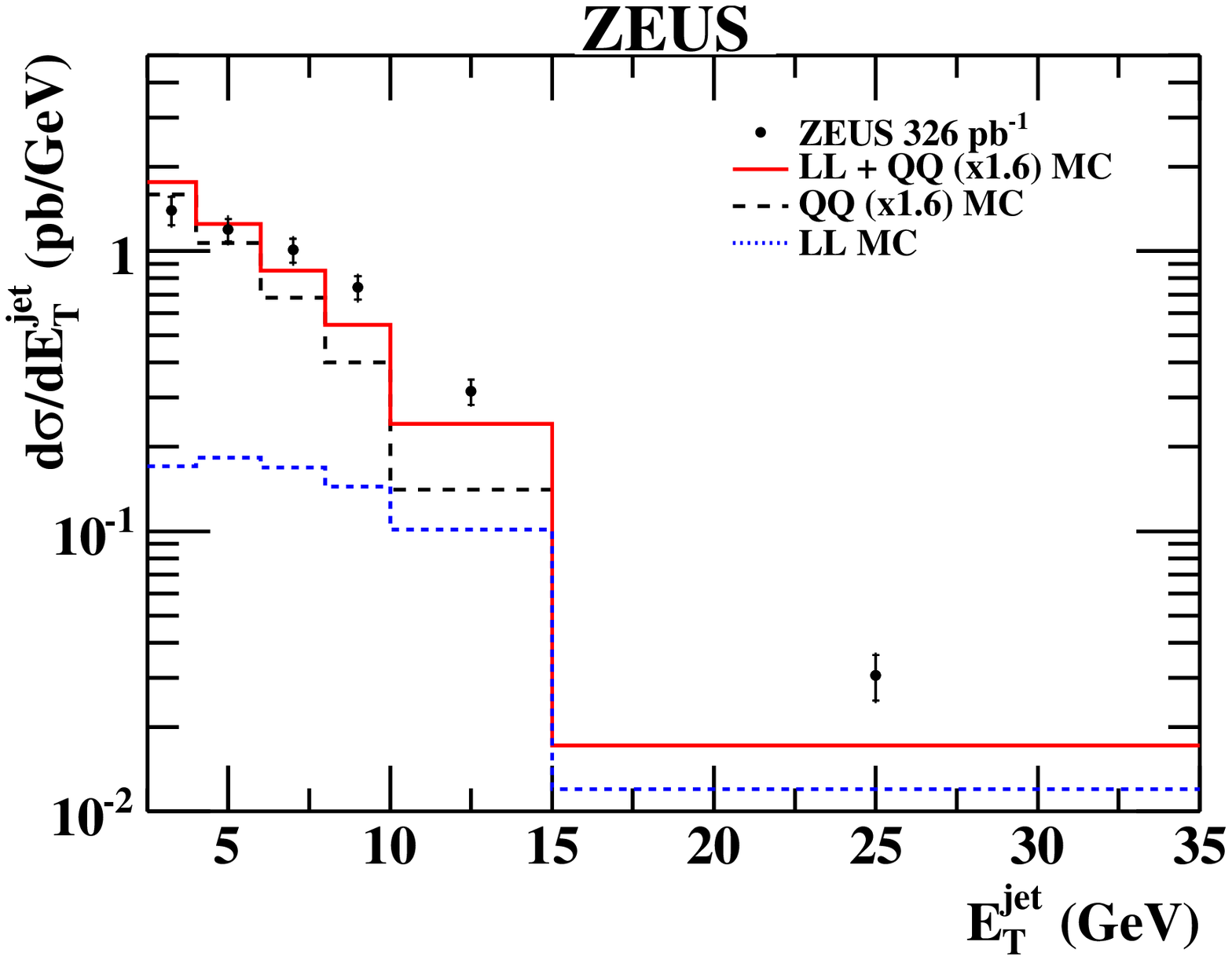,width=0.54\textwidth}
\hspace*{-0.04\textwidth}
\epsfig{file=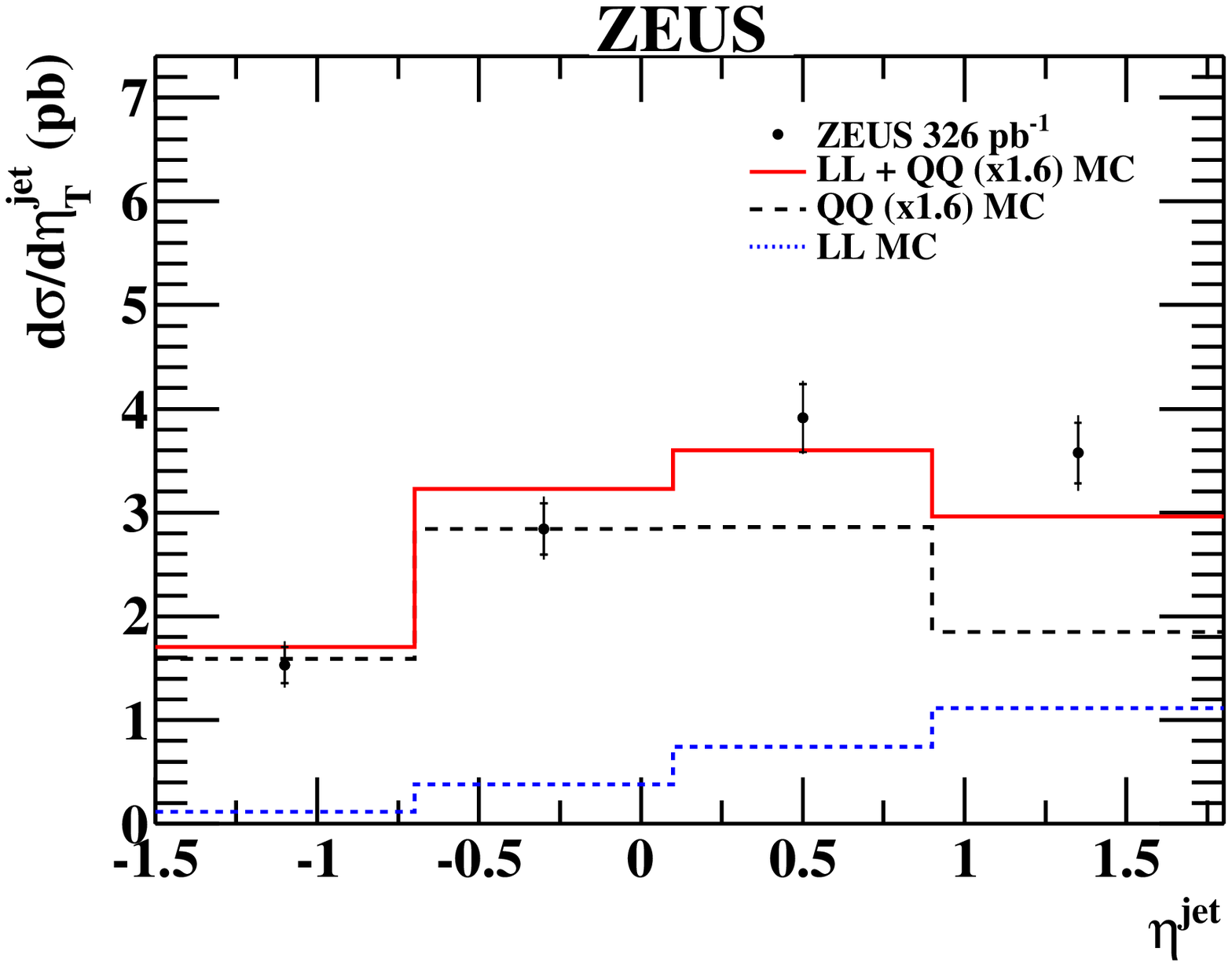,width=0.54\textwidth}
}\\[-0.04\textwidth]
\hspace*{0.1\textwidth}(e)\hspace*{0.48\textwidth}(f)\\[0.04\textwidth]
\end{center}
\caption{\small 
Isolated photon differential cross sections in 
(a) $Q^{2}$, (b) $x$, 
(c) $E_{T}^{\gamma}$,
(d) $\eta^{\gamma}$,
(e) $\ETjet$, and 
(f) $\etajet$.
The inner and outer
error bars show, respectively, the statistical uncertainty and the
statistical and systematic uncertainties added in quadrature. The
solid histograms are the reweighted Monte Carlo predictions from the sum of QQ
photons from {\sc Pythia} normalised by a factor 1.6 plus {\sc
Djangoh} LL photons. The dashed (dotted) lines show the QQ (LL)
contributions.}
\label{fig:xsec1}
\vfill
\end{figure}

The differential
cross sections as functions of $Q^2$, $x$, $E_T^{\gamma}$,
$\eta^{\gamma}$, $\ETjet$ and $\etajet$ are shown in
Fig.~\ref{fig:xsec1} and given in Tables
\ref{tab:dsdq}--\ref{tab:dsdetajet}.  As expected, the cross section
decreases with increasing $Q^2$, $x$, $E_T^{\gamma}$, and \ETjet. The
modest dependence of the cross section on $\eta^{\gamma}$ and \etajet can be
attributed to the LL contribution.  The predictions for the sum of the
expected LL contribution from {\sc Djangoh} and a factor of 1.6 times
the expected QQ contribution from {\sc Pythia} agree well with the
measurements, and this model therefore provides a good description of
the process.

\begin{figure}[p]
\vfill
\begin{center}
\mbox{
\hspace*{-0.05\textwidth}
\epsfig{file=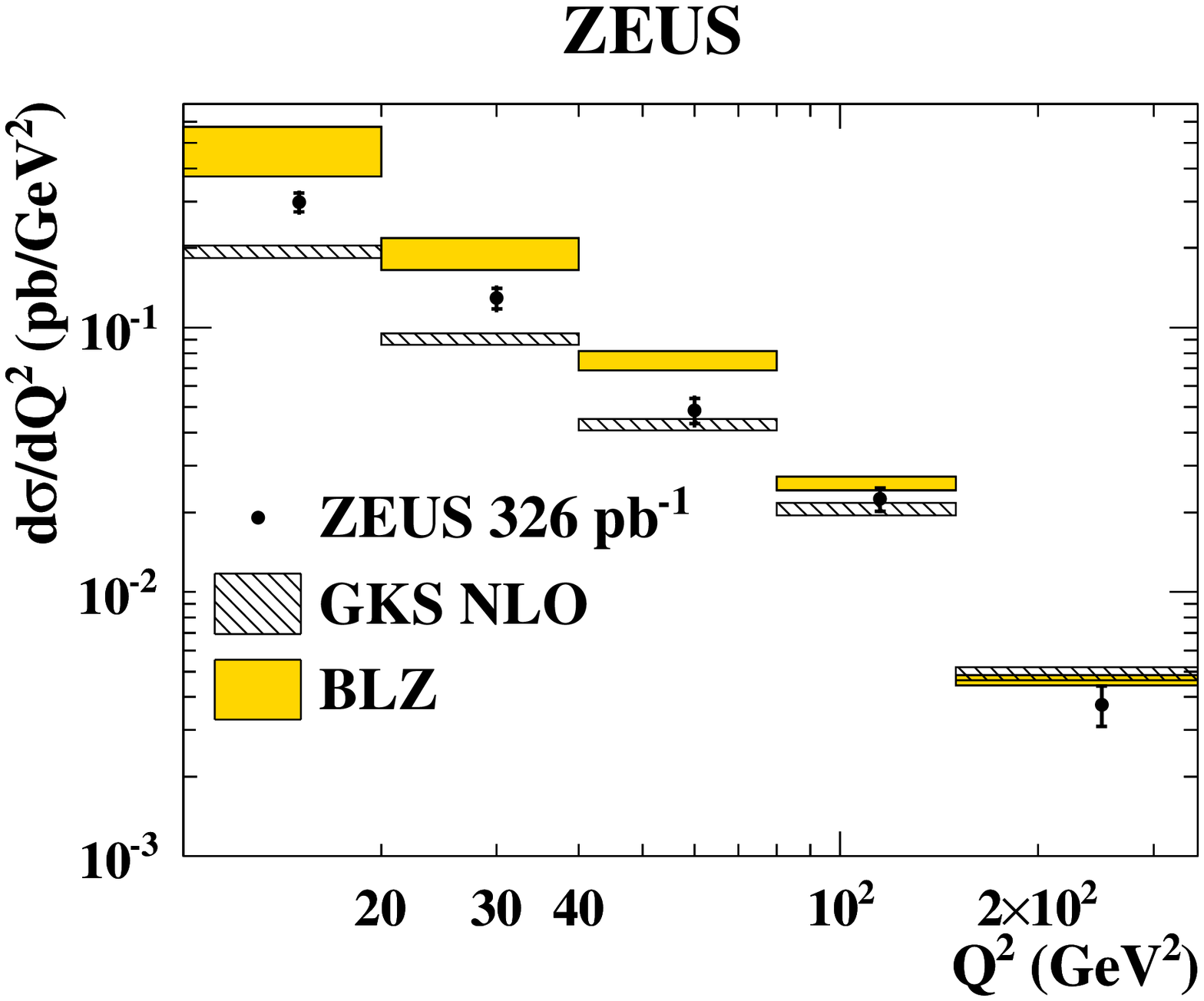,width=0.51\textwidth}
\hspace*{-0.04\textwidth}
\epsfig{file=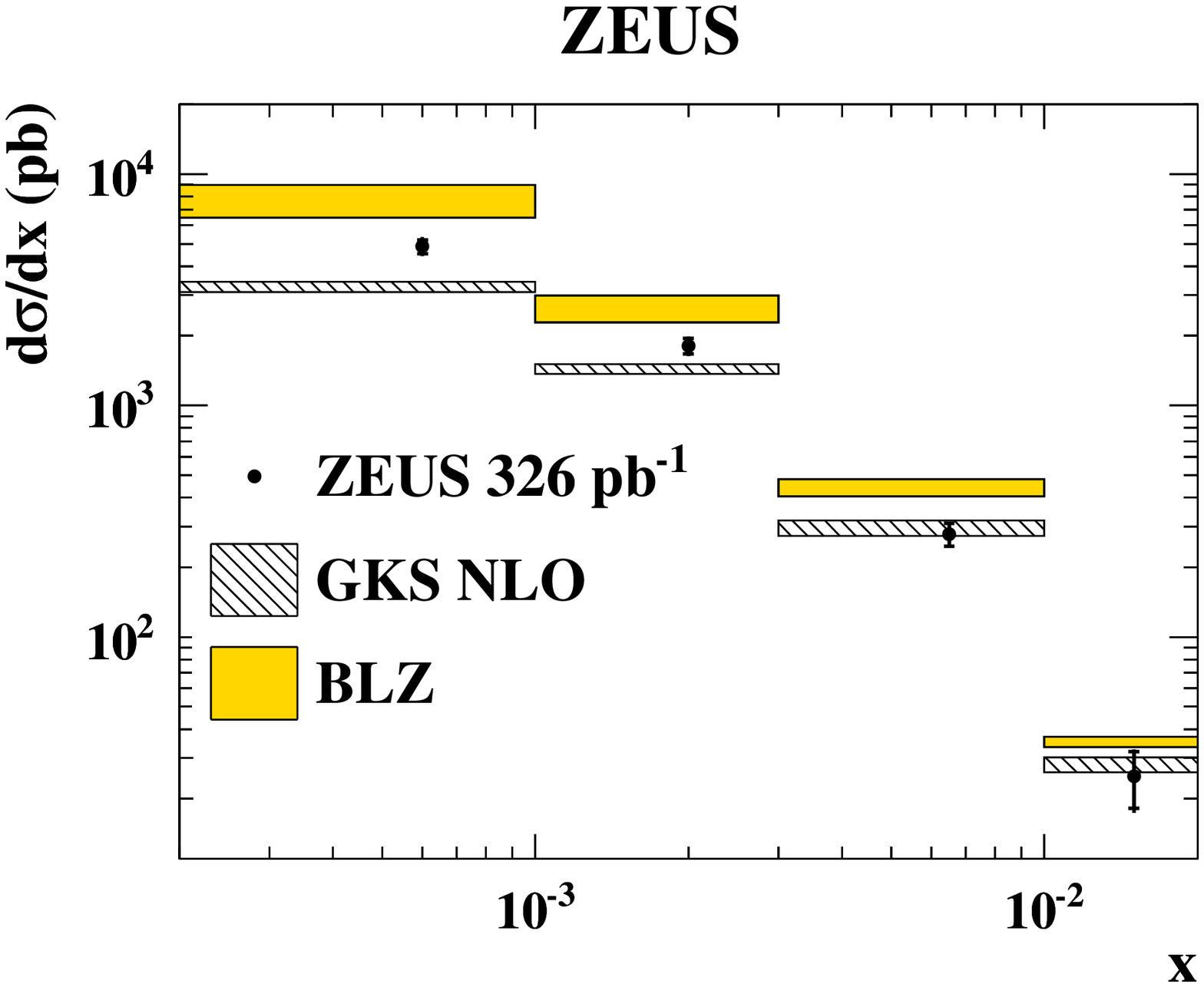,width=0.51\textwidth}
}\\[-0.04\textwidth]
\hspace*{0.1\textwidth}(a)\hspace*{0.48\textwidth}(b)\\[0.04\textwidth]
\mbox{
\hspace*{-0.05\textwidth}
\epsfig{file=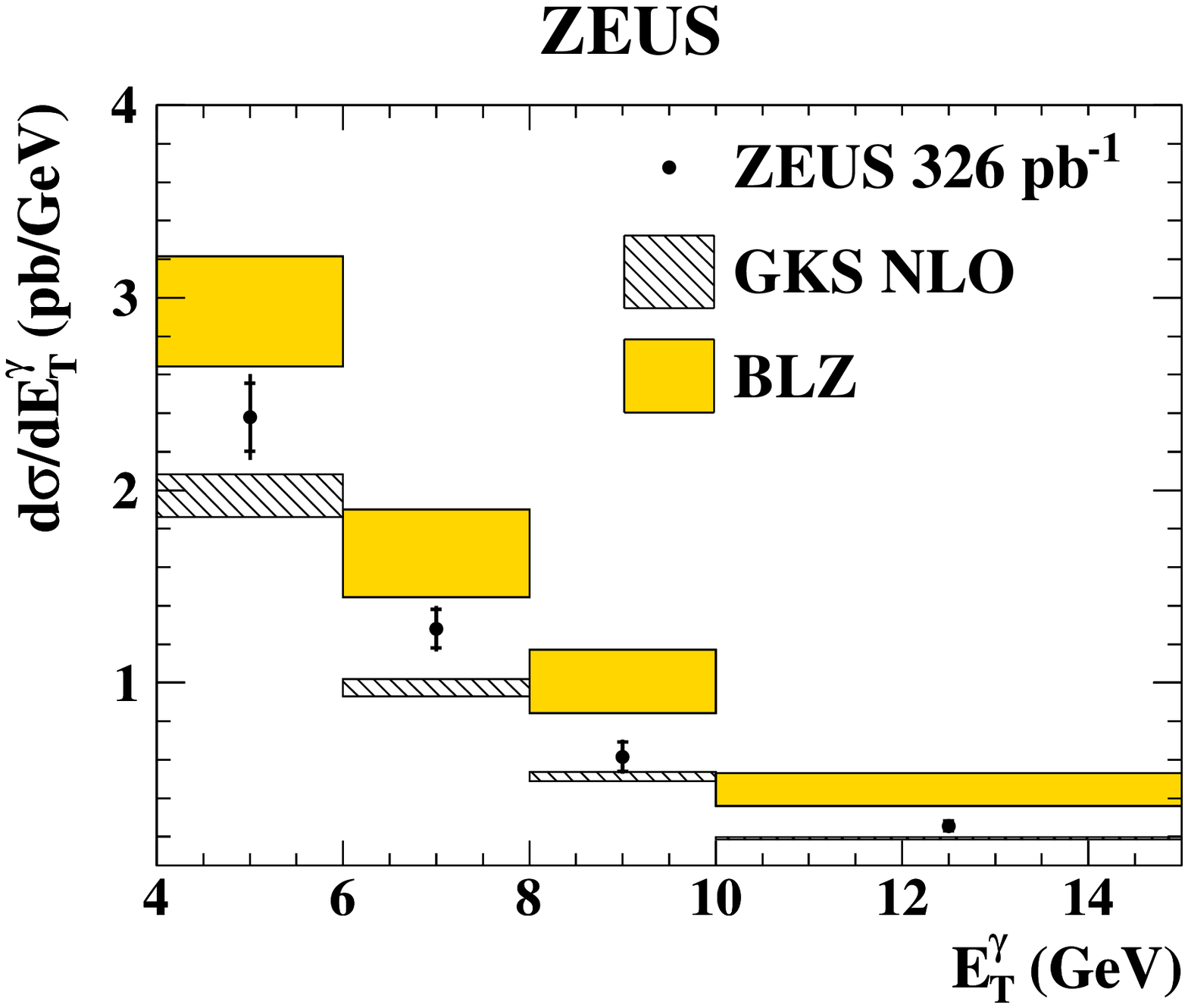,width=0.51\textwidth}
\hspace*{-0.04\textwidth}
\epsfig{file=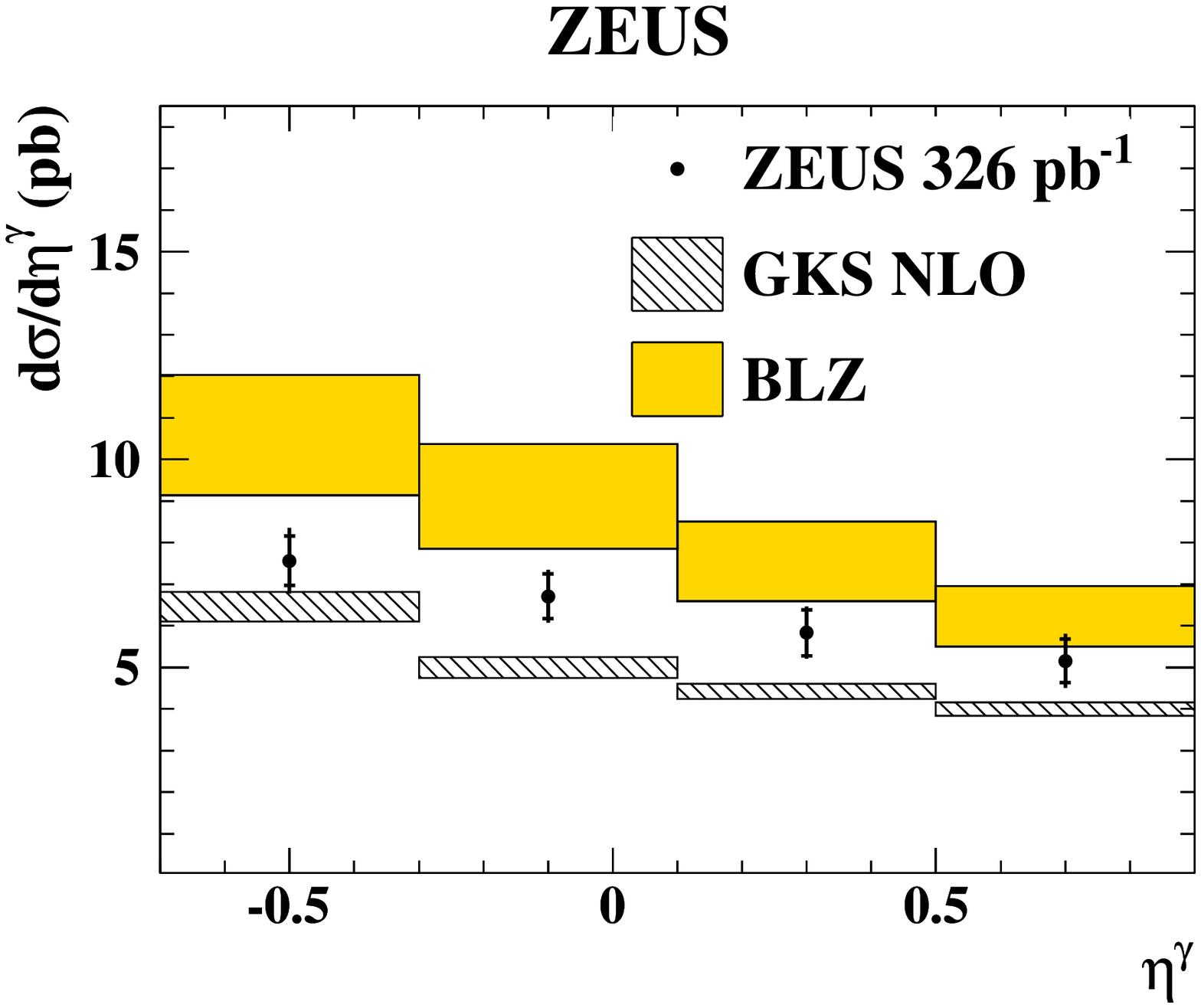,width=0.51\textwidth}
}\\[-0.04\textwidth]
\hspace*{0.1\textwidth}(c)\hspace*{0.48\textwidth}(d)\\[0.04\textwidth]
\mbox{
\hspace*{-0.05\textwidth}
\epsfig{file=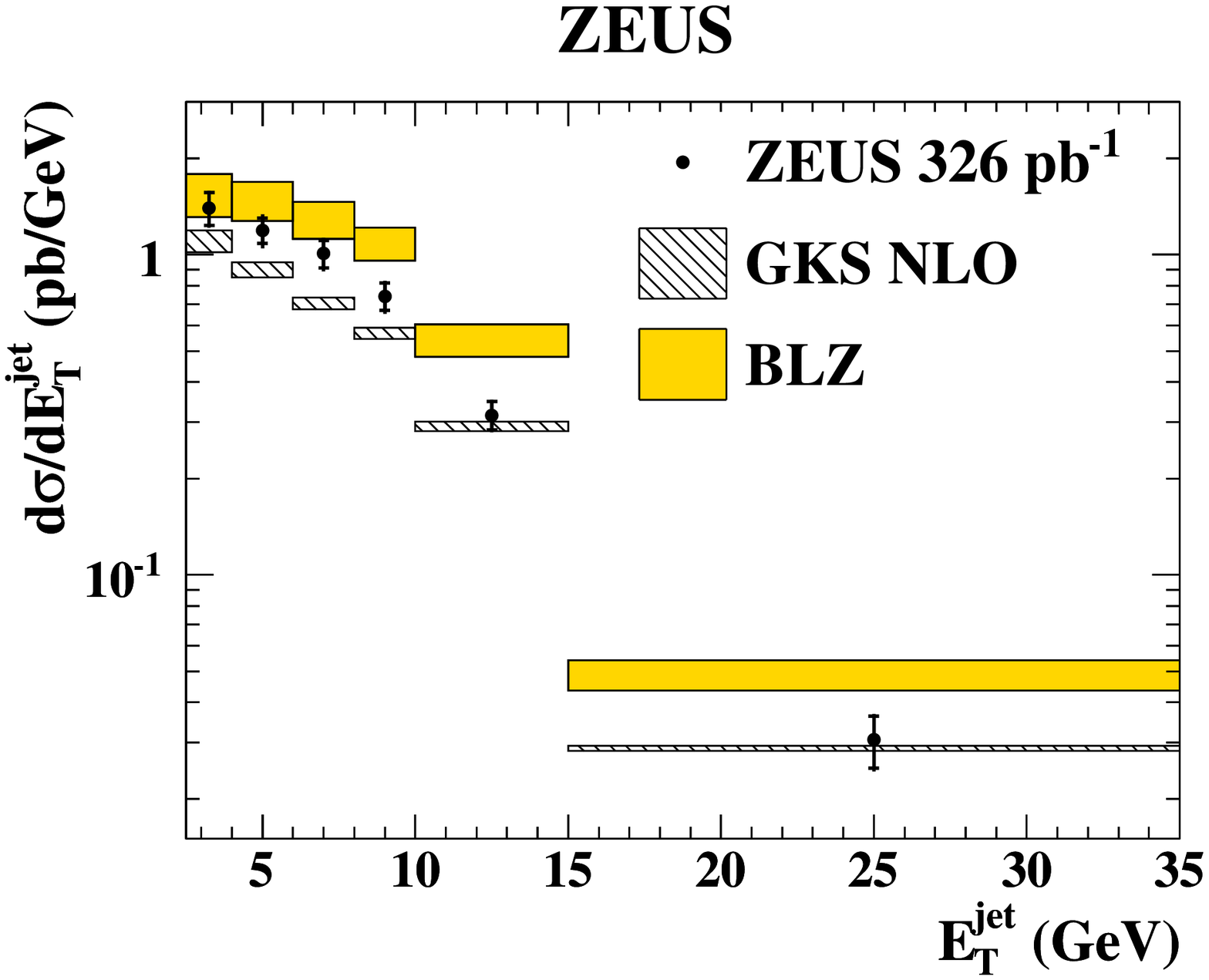,width=0.51\textwidth}
\hspace*{-0.04\textwidth}
\epsfig{file=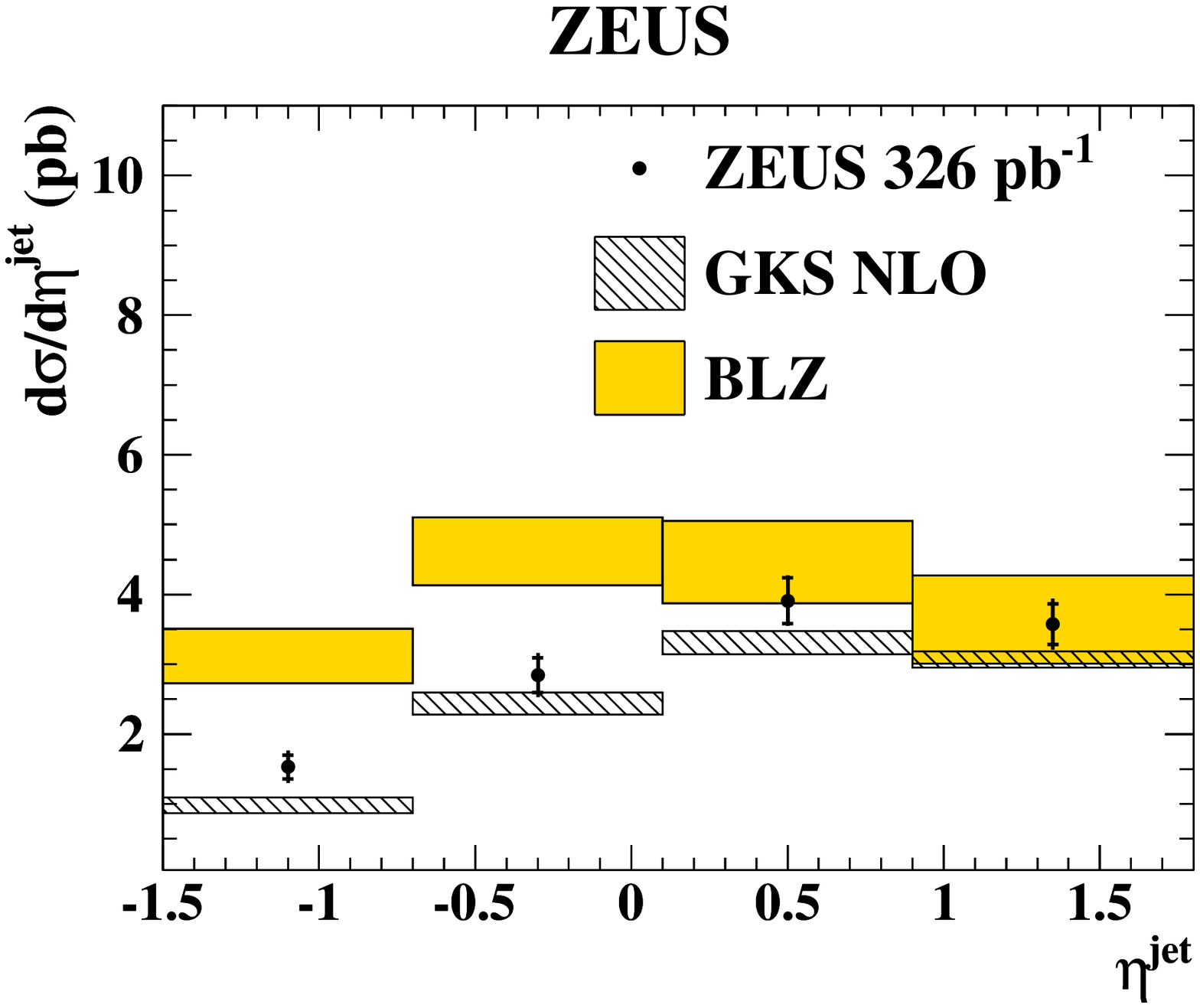,width=0.51\textwidth}
}\\[-0.04\textwidth]
\hspace*{0.1\textwidth}(e)\hspace*{0.48\textwidth}(f)\\[0.\textwidth]
\end{center}
\caption{\small 
Data points as shown in Fig.~\ref{fig:xsec1}.  Theoretical predictions from
Gehrmann-De Ridder {\it et al.}\/ (GKS)~\protect\cite{priv:spiesberger:2011}
and Baranov {\it et al.}\/ (BLZ)~\protect\cite{priv:zotov:2011} are shown,
with associated uncertainties indicated by the shaded bands.}
\label{fig:xsec3}
\vfill
\end{figure}

The theoretical predictions described in Section~\ref{sec:theory} are
compared to the measurements in Fig.~\ref{fig:xsec3}. The predictions
from GKS~\cite{priv:spiesberger:2011} describe the shape of all the
distributions reasonably well, but the rise seen at low $Q^2$ and at
low $x$ is underestimated.  The cross section as a function
of $\eta^{\gamma}$ and \etajet is underestimated by about $20 \%$. This
was also observed in the earlier inclusive photon
measurement~\cite{pl:b687:16}. The theoretical uncertainties are
indicated by the width of the shaded area.  The calculations of
BLZ~\cite{priv:zotov:2011} also describe the shape of the data
reasonably well, but the predicted overall rate is on average too high by about
20\%.

\section{Conclusions}

The production of isolated photons accompanied by jets has been
measured in deep inelastic scattering with the ZEUS detector at HERA
using an integrated luminosity of $326\,\mathrm{pb}^{-1}$.  The
present results improve on earlier ZEUS results~\cite{pl:b595:86}
which were made with an integrated luminosity of 121 pb$^{-1}$ in a
more restricted kinematic region.  Differential cross sections as
functions of several variables are presented within the kinematic
region defined by: $10 < Q^2< 350 \, \mathrm{\GeV}^2$, $E'_e >
10\,\mathrm{\GeV}$, $\theta_e > 140^{\circ}$, $-0.7< \eta^\gamma <
0.9$, $4 < E_T^{\gamma}< 15\,\mathrm{\GeV\ } $, $\ETjet > 2.5$ \GeV\
and $-1.5 <\etajet< 1.8$ in the laboratory frame.  The order
$\alpha^3\alpha_s$ predictions of Gehrmann-de Ridder {\it et al.\
}reproduce the shapes of all the measured experimental distributions
reasonably well, as do the predictions of Baranov {\it et al.}
However neither calculation gives a correct normalisation.  The
results presented here can be used to make
further improvements in the QCD calculations.

\section*{Acknowledgements}
\label{sec-ack}

\Zacknowledge\ We also thank H. Spiesberger and N. Zotov for providing theoretical results.

\vfill\eject

\providecommand{\etal}{et al.\xspace}
\providecommand{\coll}{Collaboration}
\catcode`\@=11
\def\@bibitem#1{%
\ifmc@bstsupport
  \mc@iftail{#1}%
    {;\newline\ignorespaces}%
    {\ifmc@first\else.\fi\orig@bibitem{#1}}
  \mc@firstfalse
\else
  \mc@iftail{#1}%
    {\ignorespaces}%
    {\orig@bibitem{#1}}%
\fi}%
\catcode`\@=12

\clearpage

\newcommand{\TO}{\hspace*{-1.5ex}--\hspace*{-1.5ex}}
\newcommand{\BR}{\hspace*{1.2ex}}
\newcommand{\B}{$\hspace*{1.2ex}$}
\newcommand{\sta}{\mathrm{(stat.)}}
\newcommand{\sys}{\mathrm{(sys.)}}
\renewcommand{\strut}{\rule[-1.4ex]{0ex}{4ex}}
\begin{table}
\begin{center}
\begin{tabular}{|r|ll|    }
\hline
\multicolumn{1}{|c|}{$Q^2$ range}   &  &           \\ [-1.0ex]
\multicolumn{1}{|c|}{ ($\mathrm{GeV}^{2}$)}   & \multicolumn{2}{|c|}{\raisebox{1.5ex}{{\large$\frac{d\sigma}{dQ^2}$} ($\mathrm{pb}\,\mathrm{GeV}^{-2}$)}} \\
\hline
10 -- \BR20 & $0.298$&$\pm 0.024\B\,\sta\,\pm 0.019\,\B\sys$  \\
20 -- \BR40 & $0.129$&$\pm 0.012\B\,\sta\,\pm 0.009\,\B\sys$ \\
40 -- \BR80 & $0.049$&$\pm 0.005\B\,\sta\,\pm 0.004\,\B\sys$ \\
80 -- 150 & $0.0224$&$\pm 0.0023\,\sta\,\pm 0.0011\,\sys$ \\
150 -- 350 & $0.0037$&$\pm 0.0007\,\sta\,\pm 0.0002\,\sys$ \\
\hline
\end{tabular}
\end{center}
\caption{Measured differential cross-section
$\frac{d\sigma}{dQ^2}$. The quoted systematic uncertainty includes all the
components added in quadrature. 
\label{tab:dsdq}}
\end{table}

\begin{table}
\begin{center}
\begin{tabular}{|lcl|ll|}
\hline
\multicolumn{3}{|c|}{\strut$x$ range}    & \multicolumn{2}{|c|}{ {\large $\frac{d\sigma}{dx}$} ($\mathrm{pb}$)}  \\
\hline
0.0002 &\!--\!& 0.001&$4869$ &$\pm334\,\sta\,\pm312\,\sys$\\
0.001 &\!--\!& 0.003 &$1811$ &$\pm139\,\sta\,\,\pm\B104\,\sys$ \\
0.003 &\!--\!& 0.01&$ \B278 $&$\pm\B31\,\sta\,\,\pm\B13\,\sys$\\
0.01 &\!--\!&0.02&$\B\B25$&$\pm\B\B7\,\sta\,\,\pm\B\B3\,\sys$\\
\hline
\end{tabular}
\end{center}
\caption{Measured differential cross-section $\frac{d\sigma}{dx}$. Details as in Table \ref{tab:dsdq}. \label{tab:dsdx}}
\end{table}

\begin{table}
\begin{center}
\begin{tabular}{|rcl|ll|}
\hline
\multicolumn{3}{|c|}{$E_T^\gamma$ range}   &  &           \\ [-2.0ex]
\multicolumn{3}{|c|}{(GeV)} & \multicolumn{2}{|c|}{\raisebox{2.5ex}{{\large$\frac{d\sigma}{dE^{\gamma}_T}$} ($\mathrm{pb}\,\mathrm{GeV}^{-1})$}} \\[-0.0ex] 
\hline
4 &\!--\!& 6   & $2.38$&$ \pm 0.18\,\sta\,\pm 0.13\,\sys$ \\
6 &\!--\!& 8   & $1.28$&$ \pm 0.10\,\sta\,\pm 0.06\,\sys$ \\
8 &\!--\!& 10  & $0.62$&$ \pm 0.08\,\sta\,\pm 0.04\,\sys$ \\
10 &\!--\!& 15 & $0.26$&$ \pm 0.03\,\sta\,\pm 0.02\,\sys$ \\
\hline
\end{tabular}
\end{center}
\caption{Measured differential cross-section $\frac{d\sigma}{dE^{\gamma}_T}$. Details as in Table \ref{tab:dsdq}. \label{tab:dsdet}}
\end{table}

\begin{table}
\begin{center}
\begin{tabular}{|rcr|cc|}
\hline
\multicolumn{3}{|c|}{ $\eta^{\gamma}$ range }  & \multicolumn{2}{|c|}{{\large $\frac{d\sigma}{d\eta^{\gamma}}$} ($\mathrm{pb}$)}     \\
\hline
--0.7 &\!--\!& --0.3 & $7.6$&$ \pm 0.6\,\sta\,\pm 0.5\,\sys$ \\
--0.3 &\!--\!& 0.1 & $ 6.7$&$ \pm 0.5\,\sta\,\pm 0.3\,\sys$ \\
0.1 &\!--\!& 0.5  & $ 5.8$&$ \pm 0.6\,\sta\,\pm 0.3 \,\sys$ \\
0.5 &\!--\!& 0.9 & $  5.2$&$ \pm 0.5\,\sta\,\pm 0.4 \,\sys$ \\
\hline
\end{tabular}
\end{center}
\caption{Measured differential cross-section $\frac{d\sigma}{d\eta^{\gamma}}$. Details as in Table \ref{tab:dsdq}.\label{tab:dsdeta}}
\end{table}

\begin{table}
\begin{center}
\begin{tabular}{|rcr|ll|}
\hline
\multicolumn{3}{|c|}{$\ETjet      $ range}   & &\\[-2.0ex]
\multicolumn{3}{|c|}{ \raisebox{1.5ex}{(\GeV)}}   &  \multicolumn{2}{|c|}{\raisebox{3.5ex}{{\large$\frac{d\sigma}{d\ETjet}$} ($\mathrm{pb}\,\mathrm{GeV}^{-1})$}} \\[-1ex]
\hline
2.5 &\!--\!& 4   & $1.40$&$ \pm  0.16\B\,\sta\,\pm 0.08\B\,\sys$ \\
4 &\!--\!& 6   & $1.19$&$ \pm  0.11\B\,\sta\, \pm0.10\B\,\sys$ \\
6 &\!--\!&  8  & $1.01$&$ \pm  0.10\B\,\sta\,\pm 0.07\B\,\sys$ \\
8  &\!--\!& 10 & $0.74$&$ \pm  0.07\B\,\sta\,\pm 0.05\B\,\sys$ \\
10&\!--\!& 15  & $0.32$&$ \pm  0.03\B\,\sta\,\pm 0.02\B\,\sys$ \\
15&\!--\!& 35  & $0.031$&$\pm  0.006\,\sta\,\pm 0.003\,\sys$ \\

\hline
\end{tabular}
\end{center}
\caption{Measured differential cross-section $\frac{d\sigma}{d\ETjet}$. Details as in Table \ref{tab:dsdq}.\label{tab:dsdetjet}}
\end{table}

\begin{table}
\begin{center}
\begin{tabular}{|rcr|cc|}
\hline
\multicolumn{3}{|c|}{ $\etajet      $ range }  & \multicolumn{2}{|c|}{ {\large$\frac{d\sigma}{d\etajet}$} ($\mathrm{pb}$)} \\
\hline
--1.5 &\!--\!& --0.7   & $1.53$&$ \pm 0.17\,\sta\,\pm 0.15\,\sys$ \\
--0.7 &\!--\!& 0.1   & $2.84$&$ \pm 0.25\,\sta\,\pm 0.19\,\sys$ \\
0.1 &\!--\!& 0.9  & $3.91$&$ \pm 0.33\,\sta\,\pm 0.14\,\sys$ \\
0.9 &\!--\!& 1.8 & $3.57$&$ \pm 0.29\,\sta\,\pm 0.22\,\sys$ \\
\hline
\end{tabular}
\end{center}
\caption{Measured differential cross-section $\frac{d\sigma}{d\etajet      }$. Details as in Table \ref{tab:dsdq}.\label{tab:dsdetajet}}
\end{table}

\newpage
\clearpage


\newpage

\end{document}